\documentclass[aps,pra,floatfix,showpacs,preprint]{revtex4-1}

\usepackage{graphicx}
\usepackage{amsmath, amsfonts, amssymb, bm}

\usepackage{amstext}
\usepackage{mathrsfs}
\usepackage{color}

\begin{document}
\title{High-energy behavior of strong-field QED in an intense plane wave}
\author{T.~Podszus}
\author{A.~Di Piazza}
\email{dipiazza@mpi-hd.mpg.de}
\affiliation{Max Planck Institute for Nuclear Physics, Saupfercheckweg 1, D-69117 Heidelberg, Germany}

\begin{abstract}
Analytical calculations of radiative corrections in strong-field QED have hinted that in the presence of an intense plane wave the effective coupling of the theory in the high-energy sector may increase as the $(2/3)$-power of the energy scale. These findings have raised the question of their compatibility with the corresponding logarithmic increase of radiative corrections in QED in vacuum. However, all these analytical results in strong-field QED have been obtained within the limiting case of a background constant crossed field. Starting from the polarization operator and the mass operator in a general plane wave, we show that the constant-crossed-field limit and the high-energy limit do not commute with each other and identify the physical parameter discriminating between the two alternative limits orders. As a result, we find that the power-law scaling at asymptotically large energy scales pertains strictly speaking only to the case of a constant crossed background field, whereas high-energy radiative corrections in a general plane wave depend logarithmically on the energy scale as in vacuum. However, we also confirm the possibility of testing the ``power-law'' regime experimentally by means of realistic setups involving, e.g., high-power lasers or high-density electron-positron bunches.  
\end{abstract}

\pacs{12.20.Ds, 41.60.-m}
\maketitle

\section{Introduction}
The predictions of QED agree with experiments with astonishing accuracy (see, e.g., Refs. \cite{Hanneke_2008,Sturm_2011}). The question as whether QED can be considered a truly fundamental theory, however, relates to its behavior at asymptotic high energies. Now, the QED coupling $\alpha=e^2/\hbar c$, with $e<0$ being the electron charge and in units where $4\pi\epsilon_0=1$, is about $1/137$ at ordinary energies of the order of, say, the electron rest energy $mc^2=0.511\;\text{MeV}$. At higher and higher energies, the coupling $\alpha$ increases and features a pole, called Landau pole, at $\Lambda_{\text{QED}}\sim mc^2\exp(3\pi/2\alpha)\sim 10^{277}\;\text{GeV}$ \cite{Jauch_b_1976,Itzykson_b_1980,Landau_b_4_1982,Schwartz_b_2014}. The existence of the Landau pole has profound theoretical implications and, strictly speaking, prevents one from considering QED as a fundamental theory. From a more pragmatic perspective, however, it is clear that, because of its extremely large value, the existence of the Landau pole does not represent a real limitation on the applicability of QED. Moreover, numerous experimental evidences have already called for embedding QED into a more general theory, the Standard Model, at much lower energies than $\Lambda_{\text{QED}}$. The exceedingly large value of $\Lambda_{\text{QED}}$ is intimately related to the fact that radiative corrections in QED increase logarithmically for increasing energies \cite{Jauch_b_1976,Itzykson_b_1980,Landau_b_4_1982,Schwartz_b_2014}.

The great success of QED has been motivating to test the theory under more extreme conditions as, e.g., those provided by intense background electromagnetic fields. The typical electromagnetic field scale of QED is set by the so-called ``critical'' field of QED: $F_{cr}=m^2/|e|=1.3\times 10^{16}\;\text{V/cm}=4.4\times 10^{13}\;\text{G}$ (from now on units with $\hbar=c=1$ are employed) \cite{Landau_b_4_1982,Fradkin_b_1991,Dittrich_b_1985}. The vacuum becomes unstable in the presence of an electric field of the order of $F_{cr}$ and the interaction energy of the electron magnetic moment with a magnetic field of the order of $F_{cr}$ is comparable with the electron rest energy (note that the electric field experienced by the bound electron in the experiment reported in Ref. \cite{Sturm_2011} was about $10^{-3}F_{cr}$). Generally speaking, the presence of intense background electromagnetic fields allows for testing QED on a sector where nonlinear effects with respect to the background field strongly affect physical processes and the dynamics of charged particles.

High-power optical laser facilities are a prospective tool to test QED at critical field strengths, which correspond to laser intensities of the order of $10^{29}\;\text{W/cm$^2$}$. In fact, although available lasers have reached peak intensities $I_0$ of the order of $10^{22}\;\text{W/cm$^2$}$ \cite{Yanovsky_2008} and upcoming facilities aim at $I_0\sim 10^{23}\text{-}10^{24}\;\text{W/cm$^2$}$ \cite{APOLLON_10P,ELI,CoReLS}, the Lorentz invariance of QED implies that the effective laser field strength at which a process occurs is the one experienced by the charged particles in their rest frame \cite{Mitter_1975,Ritus_1985,Ehlotzky_2009,Reiss_2009,Di_Piazza_2012,Dunne_2014}. More quantitatively, if $F_0^{\mu\nu}=(\bm{E}_0,\bm{B}_0)$ denotes a measure of the amplitude of the laser electromagnetic field and if a quantum process is initiated by an electron/positron (photon) with four-momentum $p^{\mu}=(\varepsilon,\bm{p})$ ($k^{\mu}=(\omega,\bm{k})$), the effective field strength in units of $F_{cr}$ is provided by the gauge- and Lorentz-invariant quantum nonlinearity parameter $\chi_0=\sqrt{|(F_{0,\mu\nu}p^{\nu})^2|}/mF_{cr}$ ($\kappa_0=\sqrt{|(F_{0,\mu\nu}k^{\nu})^2|}/mF_{cr}$) \cite{Mitter_1975,Ritus_1985,Ehlotzky_2009,Reiss_2009,Di_Piazza_2012,Dunne_2014}. Thus, the strong-field QED regime ($\chi_0,\kappa_0\gtrsim 1$) can be entered already at intensities of the order of $10^{23}\;\text{W/cm$^2$}$, if the laser field counterpropagates with respect to an electron/positron (photon) of energy of the order of $500\;\text{MeV}$. 

Now, electron beams with energies of the order of $50\;\text{GeV}$ have been already produced \cite{Bula_1996,Burke_1997} and one can even imagine to enter a regime of unprecedented field strengths where $\chi_0,\kappa_0\gg 1$ (see Refs. \cite{Blackburn_2018b,Baumann_2018,Yakimenko_2018} for recent proposals to enter this regime via laser-electron interaction \cite{Blackburn_2018b,Baumann_2018} and via beamstrahlung \cite{Yakimenko_2018}). The regime $\chi_0,\kappa_0\gg 1$ is theoretically extremely interesting especially due to the so-called ``Ritus-Narozhny (RN) conjecture'' \cite{Ritus_1970,Narozhny_1979,Narozhny_1980,Morozov_1981} about the high-energy behavior of radiative corrections in strong-field QED in a constant crossed field (CCF) (see also Ref. \cite{Akhmedov_1983} and the reviews in Refs. \cite{Akhmedov_2011,Fedotov_2017}). We recall that a CCF is a constant and uniform electromagnetic field $(\bm{E}_0,\bm{B}_0)$ such that the two field Lorentz-invariants $\bm{E}_0^2-\bm{B}_0^2$ and $\bm{E}_0\cdot\bm{B}_0$ vanish. The RN conjecture states that at $\chi_0\gg 1$ ($\kappa_0\gg 1$) the effective coupling of QED in a CCF scales as $\alpha\chi_0^{2/3}$ ($\alpha\kappa_0^{2/3}$). Since, apart from irrelevant prefactors, the energy of the incoming particle enters radiative corrections only through $\chi_0$ ($\kappa_0$) at $\chi_0\gg 1$ ($\kappa_0\gg 1$), the RN conjecture implies an asymptotic high-energy behavior of strong-field QED in a CCF qualitatively different from that of QED in vacuum. The physical relevance of the RN conjecture is broadened by the so-called local constant field limit, stating that in the limit of low-frequency plane waves the probabilities of QED processes reduce to the corresponding probabilities in a CCF averaged over the phase-dependent plane-wave profile \cite{Ritus_1985}.

The aim of the present work is to show that the high-energy limit and the low-frequency limit do not commute with each other and that consequently the power-law scaling of the effective coupling constant at asymptotically large energy scales strictly speaking pertains only to the CCF background field. Instead, in the case of a general plane wave the asymptotic scaling of radiative corrections at high energies is shown to be logarithmic as in vacuum. It is worth emphasizing here that, being an approximation, it is not surprising that under certain circumstances the local constant field limit may give even qualitatively different results from the exact theory in a plane-wave field. Indeed, we recall that the basic assumption behind the  local constant field approach is that the quantum process at hand is formed on a length which is much smaller than the typical laser wavelength \cite{Ritus_1985}. The analysis below shows that in the high-energy limit this assumption is violated and that radiative corrections are formed over much longer regions.

Our investigation starts from the one-loop polarization operator (see Fig. 1) and mass operator (see Fig. 2) in a general plane wave.
\begin{figure}
\begin{center}
\includegraphics[width=0.6\columnwidth]{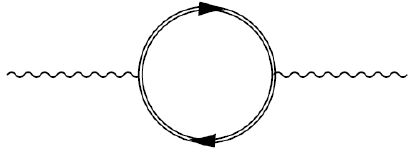}
\caption{The one-loop polarization operator in an intense plane wave. The double lines represent exact electron propagators in a plane wave (Volkov propagators) \cite{Landau_b_4_1982}.}
\end{center}
\end{figure}
\begin{figure}
\begin{center}
\includegraphics[width=0.6\columnwidth]{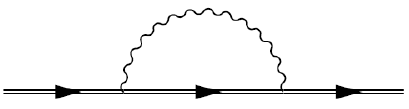}
\caption{The one-loop mass operator in an intense plane wave. The double lines represent exact electron states and propagator in a plane wave (Volkov states and propagator, respectively) \cite{Landau_b_4_1982}.}
\end{center}
\end{figure}

The one-loop polarization operator in a general plane-wave background field has been first evaluated in Refs. \cite{Becker_1975,Baier_1976_b}. However, it turned out to be technically more convenient here to employ an equivalent expression of the polarization operator found more recently in Ref. \cite{Meuren_2013}. The corresponding expression of the mass operator has been found in Ref. \cite{Baier_1976_a}. In all these works the external plane-wave field has been taken into account exactly in the calculations by employing the Furry picture \cite{Furry_1951}, i.e., by quantizing the electron-positron field starting from the Dirac Lagrangian in the presence of the background plane-wave field. This is indicated in the diagrams in Figs. 1 and 2 by representing the electron states in the plane wave (Volkov states) and the electron propagator in the plane wave (Volkov propagator) by means of double lines.

The paper is organized as follows. First, we investigate the polarization operator (Sec. II) and then we pass to the technically more complicated case of the mass operator (Sec. III). In order to make the presentation less abstract, the results are presented in the special case of a single-cycled laser pulse. This gives one also the possibility of introducing the analytical techniques and of understanding their region of applicability in a concrete case. Then, the results are generalized to the case of an arbitrary finite pulse in Sec. IV. The conclusions of the paper are presented in Sec. V. 

After this paper was submitted, related calculations on the probability of single photon emission and of photon helicity flip in a general plane wave, which are related to the imaginary part of the mass operator and of the polarization operator, respectively, appeared in Ref. \cite{Ilderton_2019}, whose conclusions are in agreement with ours.
%
%
\section{High-energy asymptotic of the one-loop polarization operator in a plane wave}
As we have mentioned in the Introduction, we start here from the general expression of the polarization operator in an arbitrary plane wave found in Ref. \cite{Meuren_2013}. In order to emphasize the difference between the CCF case and the plane-wave case, we choose here the most similar conditions to the CCF case, i.e., a linearly-polarized plane wave and, in agreement to the available results in a CCF \cite{Ritus_1970,Narozhny_1979,Narozhny_1980,Morozov_1981}, an on-shell incoming photon whose four-momentum $k_1^{\mu}$ ($k_1^2=0$) coincides with that $k_2^{\mu}$ of the outgoing photon, i.e., $k_1^{\mu}=k_2^{\mu}=k^{\mu}$. The plane wave propagates along a given direction $\bm{n}$, such that $k_0^{\mu}=(\omega_0,\bm{k}_0)=\omega_0(1,\bm{n})$ is the typical (on-shell, $k_0^2=0$) laser four-momentum, with $\omega_0$ being the central laser angular frequency (more generally, this quantity can be interpreted as the inverse of a typical time scale characterizing the plane wave). The direction $\bm{n}$ identifies a plane perpendicular to it, where we introduce two unit vectors $\bm{e}$ and $\bm{b}$ perpendicular to each other and to $\bm{n}$. By correspondingly defining the two four-vectors $e_e^{\mu}=(0,\bm{e})$ and $e_b^{\mu}=(0,\bm{b})$, it is always possible to write the laser four-potential $A_0^{\mu}(\varphi)=(0,\bm{A}_0(\varphi))$ in the form $A_0^{\mu}(\varphi)=A_0e_e^{\mu}\psi(\varphi)$, where the constant $A_0>0$ relates to the amplitude $F_0^{\mu\nu}$ of the plane wave as $F_0^{\mu\nu}=A_0(k_0^{\mu}e_e^{\nu}-k_0^{\nu}e_e^{\mu})$, where the well-behaved function $\psi(\varphi)$ of the phase $\varphi=(k_0x)$ is arbitrary except that it vanishes sufficiently fast for $\varphi\to\pm\infty$. More precise conditions on the pulse-shape functions will be given below. We only mention that we will not consider the idealized case of a monochromatic (infinitely long) plane-wave field apart that briefly at the end of Sec. IV.A.

In the case under consideration the vacuum part of the polarization operator vanishes after renormalization \cite{Landau_b_4_1982}. The field-dependent part of the polarization operator, instead, can be written in momentum space as
\begin{equation}
\label{PO^munu}
P_f^{\mu\nu}(k_1,k_2)=(2\pi)^3\delta^2(\bm{k}_{1,\perp}-\bm{k}_{2,\perp})\delta((k_0k_1)-(k_0k_2))\sum_{l=e,b}P_l(k)\Lambda_l^{\mu}(k)\Lambda_l^{\nu}(k),
\end{equation}
where we have extracted the usual light-cone delta-functions enforcing the conservation of three components of the four-momenta for a process occurring in a plane wave depending on $\varphi=(k_0x)$ and where $\Lambda_l^{\mu}(k)=(k_0^{\mu}e^{\nu}_l-k_0^{\nu}e^{\mu}_l)k_{\nu}/(k_0k)$ \cite{Baier_1976_b,Meuren_2013}. Note that an additional contribution to the polarization tensor $P_f^{\mu\nu}(k_1,k_2)$ has been ignored because in the case of on-shell incoming and outgoing photons ($k_1^2=k_2^2=0$) it turns out to be proportional to $k_1^{\mu}k_2^{\nu}$ and, due to gauge invariance, would not contribute to any physical amplitude \cite{Meuren_2013}. The scalar coefficients $P_l(k)$ can be written in the form \cite{Baier_1976_b,Meuren_2013}
\begin{align}
\label{P_e}
\begin{split}
P_e(k)=&-\frac{\alpha}{2\pi}m^2\int_{-\infty}^{\infty}d\varphi\int_0^{\infty}\frac{d\tau}{\tau}\int_1^{\infty}\frac{d\rho}{\rho^{3/2}}\frac{1}{\sqrt{\rho-1}}\bigg\langle 2\xi_0^2[X(\varphi,\tau)+\rho Z(\varphi,\tau)]e^{-i\frac{4\tau\rho}{\theta_0}[1+\xi_0^2Q^2(\varphi,\tau)]}\\
&\left.-i\frac{\theta_0}{4\tau\rho}\left\{e^{-i\frac{4\tau\rho}{\theta_0}[1+\xi_0^2Q^2(\varphi,\tau)]}-e^{-i\frac{4\tau\rho}{\theta_0}}\right\}\right\rangle,
\end{split}\\
\label{P_b}
\begin{split}
P_b(k)=&-\frac{\alpha}{2\pi}m^2\int_{-\infty}^{\infty}d\varphi\int_0^{\infty}\frac{d\tau}{\tau}\int_1^{\infty}\frac{d\rho}{\rho^{3/2}}\frac{1}{\sqrt{\rho-1}}\bigg\langle 2\xi_0^2\rho Z(\varphi,\tau)e^{-i\frac{4\tau\rho}{\theta_0}[1+\xi_0^2Q^2(\varphi,\tau)]}\\
&\left.-i\frac{\theta_0}{4\tau\rho}\left\{e^{-i\frac{4\tau\rho}{\theta_0}[1+\xi_0^2Q^2(\varphi,\tau)]}-e^{-i\frac{4\tau\rho}{\theta_0}}\right\}\right\rangle,
\end{split}
\end{align}
where $\theta_0=(k_0k)/m^2=(k_0+k)^2/2m^2\ge 0$ is twice the square of the total energy of the incoming photon and of a laser photon in their center-of-momentum system in units of $m^2$, where $\xi_0=|e|A_0/m$ is the classical nonlinearity parameter \cite{Ritus_1985,Di_Piazza_2012}, and where
\begin{align}
\label{X}
X(\varphi,\tau)&=\left[\frac{1}{2\tau}\int_{-\tau}^{\tau}d\tau'\psi(\varphi-\tau')-\psi(\varphi-\tau)\right]\left[\frac{1}{2\tau}\int_{-\tau}^{\tau}d\tau'\psi(\varphi-\tau')-\psi(\varphi+\tau)\right],\\
\label{Z}
Z(\varphi,\tau)&=\frac{1}{2}\left[\psi(\varphi+\tau)-\psi(\varphi-\tau)\right]^2,\\
\label{Q^2}
Q^2(\varphi,\tau)&=\frac{1}{2\tau}\int_{-\tau}^{\tau}d\tau'\psi^2(\varphi-\tau')-\frac{1}{4\tau^2}\left[\int_{-\tau}^{\tau}d\tau'\psi(\varphi-\tau')\right]^2.
\end{align}
It is worth observing at this point that the structure of the coefficient corresponding to the additional term in the polarization operator mentioned above and to the others arising from considering a more general laser polarization is similar to those in Eqs. (\ref{P_e}) and (\ref{P_b}), and their inclusion would not change the conclusions below.

The introduction of the two important gauge- and Lorentz-invariant parameters $\theta_0$ and $\xi_0$ (note that $\kappa_0=\theta_0\xi_0$) allows us to quantitatively define the low-frequency or CCF limit and the high-energy limit, and to ascertain, in particular, their commutativity. 

We first consider the low-frequency/CCF limit, which physically has to correspond to keeping the laser field amplitude and the external photon energy fixed and finite. This is realized in a Lorentz invariant way via the double limit $\xi_0\to\infty$ and $\theta_0\to 0$ such that $\kappa_0=\theta_0\xi_0$, remains fixed and finite. As it has been shown in Ref. \cite{Meuren_2013}, the expressions in Eqs. (\ref{P_e}) and (\ref{P_b}), indeed reduce to the integrals over $\varphi$ of the corresponding coefficients of the polarization operator in a CCF \cite{Ritus_1972}, with the local expression of the quantum nonlinearity parameter being given by $\kappa(\varphi)=\kappa_0|\psi'(\varphi)|$ (here and below, a primed function indicates the derivative with respect to its argument). In fact, in the limit $\xi_0\to \infty$ and $\kappa_0$ constant the phases in the coefficients $P_e(k)$ and $P_b(k)$ become very large and the main contribution to the integral in $\tau$ comes from the region $\tau\sim 1/\xi_0\ll 1$ close to the origin. This allows one to appropriately expand the functions $X(\varphi,\tau)$, $Z(\varphi,\tau)$, and $Q^2(\varphi,\tau)$ for small values of $\tau$. The resulting integral in $\tau$ can be represented in terms of Airy and Scorer functions $\text{Ai}(\cdot)$ and $\text{Gi}(\cdot)$ \cite{NIST_b_2010} and the coefficients $P_e(k)$ and $P_b(k)$ become (see the original Ref. \cite{Narozhny_1969} although the expressions below are taken from Ref. \cite{Meuren_2013}):
\begin{align}
\label{P_e_CCF}
P_{e,\text{CCF}}(k)=&-\frac{\alpha}{3\pi}m^2\int_{-\infty}^{\infty}d\varphi\int_1^{\infty}\frac{d\rho}{\rho^{3/2}}\frac{4\rho-1}{\sqrt{\rho-1}}g\left(\frac{4\rho}{\kappa(\varphi)}\right),\\
\label{P_b_CCF}
P_{b,\text{CCF}}(k)=&-\frac{\alpha}{3\pi}m^2\int_{-\infty}^{\infty}d\varphi\int_1^{\infty}\frac{d\rho}{\rho^{3/2}}\frac{4\rho+2}{\sqrt{\rho-1}}g\left(\frac{4\rho}{\kappa(\varphi)}\right),
\end{align}
where $g(z)=z^{-2/3}df(z)/dz$ and (see, e.g. Ref. \cite{NIST_b_2010})
\begin{equation}
\label{f}
f(z)=i\int_0^{\infty}dt\,e^{-i(tz+t^3/3)}=\pi[\text{Gi}(z)+i\text{Ai}(z)].
\end{equation}
If one then performs the limit $\kappa_0\to \infty$ in Eqs. (\ref{P_e_CCF}) and (\ref{P_b_CCF}), by exploiting the asymptotic properties of the function $g(\cdot)$, one finds indeed that both $P_{e,\text{CCF}}(k)$ and $P_{b,\text{CCF}}(k)$ scale as $\int d\varphi\,\kappa^{2/3}(\varphi)$. Now, as mentioned, the above procedure works if the phases in the integrands of the coefficients $P_e(k)$ and $P_b(k)$ become very large and this requires that the parameter $r_0=\xi_0^2/\theta_0$ is much larger than unity [see Eqs. (\ref{P_e}) and (\ref{P_b})]. In other words, under the CCF limit of the polarization operator one implicitly assumes that $r_0\gg 1$. A similar observation has been already made in Ref. \cite{Di_Piazza_2007} in the case of photon splitting in a plane wave and in Refs.  \cite{Baier_1989,Dinu_2016} in the case of nonlinear Compton scattering (in this latter case the validity condition of the CCF needs to be modified at low light-cone energies of the emitted photon \cite{Di_Piazza_2018c}).

We now turn to the high-energy limit, which physically has to correspond to an incoming photon with higher and higher energy colliding with a laser field with given properties. This limit is realized in a Lorentz invariant way via the double limit $\theta_0\to \infty$ and $\kappa_0\to\infty$ such that the invariant field amplitude $\xi_0=\kappa_0/\theta_0$ remains fixed and finite. This situation is quite complementary to the CCF limit because now the phases in the integrands of the coefficients $P_e(k)$ and $P_b(k)$ tend to become much smaller than unity, with the result that the integral in $\tau$ receives a substantial contribution also for large values of $\tau$. This remark and an inspection at the phases in Eqs. (\ref{P_e}) and (\ref{P_b}) imply that, unlike in the CCF limit, the parameter $r_0$ is much smaller than unity in the high-energy limit. Below, we will show that correspondingly the asymptotic behavior of the coefficients $P_e(k)$ and $P_b(k)$ is completely different from that within the CCF limit at $\kappa_0\to\infty$.

The above analysis shows that the quantity $r_0=\xi_0^2/\theta_0$ is precisely the parameter discriminating between the CCF limit ($r_0\gg 1$) and the high-energy limit ($r_0\ll 1$), which also clarifies why the two limits do not commute. Furthermore, this implies that from a physical point of view it would be more appropriate to identify the limit $\kappa_0\to\infty$ within the CCF limit as the ``high-field limit'', as it can be realized asymptotically for higher and higher laser field strengths.

We pass now to analyze explicitly the asymptotic form of the coefficients $P_e(k)$ and $P_b(k)$ in the high-energy limit $\theta_0\to \infty$ at fixed $\xi_0$. From Eqs. (\ref{P_e}) and (\ref{P_b}) it is clear that it is sufficient to consider the coefficient $P_e(k)$. We first observe that all integrals in $\rho$ can be taken analytically because they have the form
\begin{equation}
I_n=\int_1^{\infty}\frac{d\rho}{\rho^{3/2-n}}\frac{e^{-ia\rho}}{\sqrt{\rho-1}},
\end{equation}
with $n=-1,0,+1$ and $\text{Im}[a]<0$ (recall that the prescription $m^2\to m^2-i0$ is always understood \cite{Meuren_2013}). The results are \cite{Gradshteyn_b_2000}
\begin{align}
I_{-1}(a)=&\sqrt{i\pi a}e^{-ia/2}\text{W}_{-1,1}(ia),\\
I_0(a)=&\sqrt{\pi}e^{-ia/2}\text{W}_{-1/2,1/2}(ia),\\
\label{I_1}
I_1(a)=&e^{-ia/2}\text{K}_0\left(\frac{ia}{2}\right),
\end{align}
where $\text{W}_{p,q}(\cdot)$ is the Whittaker function and $\text{K}_p(\cdot)$ is the modified Bessel function \cite{NIST_b_2010}. In our case, it is either $a=a_0(\tau)=4\tau/\theta_0$ or $a=a_0(\tau)+a_f(\varphi,\tau)$, with $a_f(\varphi,\tau)=4\tau\xi_0^2Q^2(\varphi,\tau)/\theta_0$:
\begin{equation}
\label{P_e_1}
\begin{split}
&P_e(k)=-\frac{\alpha m^2}{2\pi}\int_{-\infty}^{\infty}d\varphi\int_0^{\infty}\frac{d\tau}{\tau}\bigg\{2\xi_0^2[X(\varphi,\tau)\\
&\times I_0(a_0(\tau)+a_f(\varphi,\tau))+Z(\varphi,\tau)I_1(a_0(\tau)+a_f(\varphi,\tau))]\\
&\left.-i\frac{\theta_0}{4\tau}[I_{-1}(a_0(\tau)+a_f(\varphi,\tau))-I_{-1}(a_0(\tau))]\right\}.
\end{split}
\end{equation}
In order to be able to take the integral in $\varphi$ explicitly, the pulse-shape function $\psi(\varphi)$ has to be assigned. Since we would like to consider the more realistic case of a pulsed field than a monochromatic plane wave, for the sake of definiteness we choose the one-cycle, pulsed function $\psi(\varphi)=-\sinh(\varphi)/\cosh^2(\varphi)$ \cite{Mackenroth_2011}. This is a very convenient prototype of finite pulses because a single function encodes both the oscillation and the damping at $\varphi\to\pm\infty$ of the field and the case of a general pulsed field will be considered in Sec. IV. In this way, all the resulting integrals in the functions $X(\varphi,\tau)$, $Z(\varphi,\tau)$, and $Q^2(\varphi,\tau)$ can be taken analytically and, for the sake of convenience, we report their expressions:
\begin{align}
\label{I_X}
\mathcal{I}_X(\tau)&=\int_{-\infty}^{\infty}d\varphi\,X(\varphi,\tau)=\frac{1}{\tau^2}-2\tau\frac{3+\cosh(4\tau)}{\sinh^3(2\tau)},\\
\mathcal{I}_Z(\tau)&=\int_{-\infty}^{\infty}d\varphi\,Z(\varphi,\tau)=\frac{2}{3}+\frac{\tau\coth(\tau)-1}{\sinh^2(\tau)}+\frac{\tau\tanh(\tau)-1}{\cosh^2(\tau)},\\
\label{I_Q^2}
\mathcal{I}_{Q^2}(\tau)&=\int_{-\infty}^{\infty}d\varphi\,Q^2(\varphi,\tau)=\frac{2}{3}-\frac{1}{\tau^2}+\frac{2}{\tau}\frac{1}{\sinh(2\tau)}.
\end{align}
In particular, the function $a_f(\varphi,\tau)$ for a finite pulse, i.e., such that the pulse-shape function $\psi(\varphi)$ is square-integrable (see also Sec. IV), is bound for all values of $\tau$ [see also Eq. (\ref{Q^2})]. This suggests that in the high-energy limit $\theta_0\to \infty$, one can expand the functions $I_n(a_0(\tau)+a_f(\varphi,\tau))$ for small $a_f(\varphi,\tau)$. It is interesting to notice, as we have also hinted above, that, since the nonlinear dependence of the coefficient(s) $P_e(k)$ (and $P_b(k)$) on $\xi_0^2$ only arises through the function $a_f(\varphi,\tau)$, which is proportional to $\xi_0^2$, the high-energy limit $\theta_0\to\infty$ ultimately corresponds to the perturbative limit $\xi_0\to 0$. As we will see below, however, nonlinear effects in $\xi_0^2$ are only logarithmically suppressed as compared to the terms proportional to $\xi_0^2$. In fact, it is instructive to first consider the leading contribution in this expansion, i.e., to replace $I_1(a_0(\tau)+a_f(\varphi,\tau))\approx I_1(a_0(\tau))$, $I_0(a_0(\tau)+a_f(\varphi,\tau))\approx I_0(a_0(\tau))$ and $I_{-1}(a_0(\tau)+a_f(\varphi,\tau))-I_{-1}(a_0(\tau))\approx I'_{-1}(a_0(\tau))a_f(\varphi,\tau)$ in Eq. (\ref{P_e_1}). As we have already hinted above, all the integrals in $\varphi$ can be taken analytically and we obtain
\begin{equation}
\begin{split}
P_e(k)=&-\frac{\alpha m^2\xi_0^2}{\pi}\int_0^{\infty}\frac{d\tau}{\tau}e^{-2i\tau/\theta_0}\bigg\{\text{K}_0\left(\frac{2i\tau}{\theta_0}\right)\mathcal{I}_Z(\tau)\\
&\left.+\sqrt{\pi}\text{W}_{-1/2,1/2}\left(\frac{4i\tau}{\theta_0}\right)\left[\mathcal{I}_X(\tau)-\frac{1}{2}\mathcal{I}_{Q^2}(\tau)\right]\right\},
\end{split}
\end{equation}
which, being proportional to $\xi_0^2$, coincides with the leading-order expression of the perturbative limit $\xi_0\to 0$. Now, we evaluate the above integral in $\tau$ in the asymptotic limit $\theta_0\to\infty$. It is clear that we can divide the computation into three integrals that, according to the notation above, we will denote as $\mathscr{I}_X$, $\mathscr{I}_Z$, $\mathscr{I}_{Q^2}$:
\begin{align}
\label{int_I_X}
\mathscr{I}_X=&\sqrt{\pi}\int_0^{\infty}\frac{d\tau}{\tau}e^{-2i\tau/\theta_0}\text{W}_{-1/2,1/2}\left(\frac{4i\tau}{\theta_0}\right)\mathcal{I}_X(\tau),\\
\mathscr{I}_Z=&\int_0^{\infty}\frac{d\tau}{\tau}e^{-2i\tau/\theta_0}\text{K}_0\left(\frac{2i\tau}{\theta_0}\right)\mathcal{I}_Z(\tau),\\
\label{int_I_Q^2}
\mathscr{I}_{Q^2}=&\int_0^{\infty}\frac{d\tau}{\tau}e^{-2i\tau/\theta_0}\text{W}_{-1/2,1/2}\left(\frac{4i\tau}{\theta_0}\right)\mathcal{I}_{Q^2}(\tau).
\end{align}
The simplest integral to evaluate is $\mathscr{I}_X$ because it converges also in the limit $\theta_0\to\infty$ and its asymptotic value is $\mathscr{I}_X\approx -2/3$. The asymptotic values of the integrals $\mathscr{I}_Z$ and $\mathscr{I}_{Q^2}$ can be obtained by employing the standard technique of dividing the integration region into two regions by means of a fixed $\tau_0$ such that $1\ll\tau_0\ll \theta_0$ \cite{Bender_b_1999}. In this way, in the integrals from $0$ to $\tau_0$, the functions of $\tau/\theta_0$ in the integrands can be approximated for small values of $\tau/\theta_0$. Analogously, in the integrals from $\tau_0$ to $\infty$, the functions $\mathcal{I}_Z(\tau)$ and $\mathcal{I}_{Q^2}(\tau)$ can be approximated for large values of $\tau$. The results are
\begin{widetext}
\begin{align}
\begin{split}
\mathscr{I}_Z=&\frac{1}{3}\log^2(\theta_0)-\left(\frac{2}{3}\gamma+i\frac{\pi}{3}+C_{Z,1}\right)\log(\theta_0)+\frac{1}{3}\left(\gamma^2+i\pi\gamma-\frac{5}{12}\pi^2-\log^22\right)\\
&+\left(\gamma+i\frac{\pi}{2}\right)C_{Z,1}+\frac{1}{2}C_{Z,2}+\frac{2}{3}C_K,
\end{split}\\
\mathscr{I}_{Q^2}=&\frac{4}{3\sqrt{\pi}}\left[\log(\theta_0)-2-\gamma-i\frac{\pi}{2}\right]-\frac{2}{\sqrt{\pi}}C_{Q^2,1},
\end{align}
\end{widetext}
where $\gamma=0.577\ldots$ is the Euler constant and where
\begin{align}
\label{C_Z_1}
C_{Z,1}&=\int_0^{\infty}d\tau\log(\tau)\mathcal{I}_Z'(\tau)\approx -0.781\ldots,\\
\label{C_Z_2}
C_{Z,2}&=\int_0^{\infty}d\tau\log^2(\tau)\mathcal{I}_Z'(\tau)\approx 0.579\ldots,\\
C_K&=\int_0^{\infty}\frac{d\tau}{\tau}e^{-\tau}\left[\text{K}_0(\tau)+\gamma+\log\left(\frac{\tau}{2}\right)\right]\approx 0.240\ldots,\\
\label{C_Q^2}
C_{Q^2,1}&=\int_0^{\infty}d\tau\log(\tau)\mathcal{I}_{Q^2}'(\tau)\approx 0.218\ldots.
\end{align}

At this point the last task is to verify whether the higher-order terms in the expansion of the functions $I_n(a_0(\tau)+a_f(\varphi,\tau))$ for small $a_f(\varphi,\tau)$ provide subleading order contributions in $\theta_0$. In fact, we show explicitly that this is not the case for the function $I_1(a_0(\tau)+a_f(\varphi,\tau))$. If we expand the function $I_1(a_0(\tau)+a_f(\varphi,\tau))$ we have that the series $\tilde{\mathscr{I}}_Z$ of higher-order terms can be written as
\begin{equation}
\label{tilde_I_Z}
\tilde{\mathscr{I}}_Z=\int_0^{\infty}\frac{d\tau}{\tau}\sum_{n=1}^{\infty}\frac{\xi_0^{2n}}{n!}\left(\frac{4\tau}{\theta_0}\right)^n\left.\frac{d^nI_1(a)}{da^n}\right\vert_{a=4\tau/\theta_0}\mathcal{I}_{Z,n}(\tau),
\end{equation}
where
\begin{equation}
\label{I_Z_n}
\mathcal{I}_{Z,n}(\tau)=\int_{-\infty}^{\infty}d\varphi Z(\varphi,\tau)Q^{2n}(\varphi,\tau).
\end{equation}
The functions $\mathcal{I}_{Z,n}(\tau)$ contain neither parameters nor large numerical coefficients and tend to zero both for $\tau\to 0$ and for $\tau\to\infty$. As a result, they are different from zero only for $\tau\lesssim 1$, which can be also easily ascertained numerically. Thus, in the limit $\theta_0\to\infty$ the remaining function of $4\tau/\theta_0$ can be expanded for small values of this quantity. Since it is $d^nI_1(a)/da^n\sim (-1)^n(n-1)!/a^n$ for $a\ll 1$ and $n\ge 1$ [see Eq. (\ref{I_1}) and Ref. \cite{NIST_b_2010}], we obtain that the contribution of $\tilde{\mathscr{I}}_Z$ is independent of $\theta_0$ and equal to
\begin{equation}
\label{tI_Z}
\tilde{\mathscr{I}}_Z\equiv\tilde{\mathscr{I}}_Z(\xi_0)=-\int_{-\infty}^{\infty}d\varphi \int_0^{\infty}\frac{d\tau}{\tau}Z(\varphi,\tau)\log[1+\xi_0^2Q^2(\varphi,\tau)].
\end{equation}
Note that the definition of $Q^2(\varphi,\tau)$ in Eq. (\ref{Q^2}) implies that $Q^2(\varphi,\tau)\ge 0$. A similar analysis shows that the higher-order terms arising from the expansion of the functions $I_{-1,0}(a_0(\tau)+a_f(\varphi,\tau))$ for small $a_f(\varphi,\tau)$ are indeed subleading in $\theta_0$. Thus, we obtain
\begin{align}
\label{P_e_f}
\begin{split}
P_e(k)=&-\frac{\alpha m^2\xi_0^2}{3\pi}\left[\log^2(\theta_0)-\left(2\gamma+i\pi+2+3C_{Z,1}\right)\log(\theta_0)+\gamma^2+i\pi\gamma-\frac{5}{12}\pi^2-\log^22+2\right.\\
&\left.+3\tilde{\mathscr{I}}_Z(\xi_0)+\left(\gamma+i\frac{\pi}{2}\right)\left(3C_{Z,1}+2\right)+\frac{3}{2}C_{Z,2}+2C_K+3C_{Q^2,1}\right],
\end{split}\\
\label{P_b_f}
\begin{split}
P_b(k)=&P_e(k)-\frac{2\alpha m^2\xi_0^2}{3\pi}.
\end{split}
\end{align}
These expressions let us to conclude that the polarization operator features a leading double-logarithmic behavior in the high-energy limit. Also, the dependence on the classical nonlinearity parameter is quadratic except that for the function $\tilde{\mathscr{I}}_Z(\xi_0)$, which depends on the logarithm of the ratio between the local value of the square of the electron laser-dressed mass and $m^2$ [see Eqs. (\ref{tI_Z}) and (\ref{Q^2})], and contributes to the constant term in the asymptotics. On the one hand, this is certainly different from the corresponding vacuum case, as the polarization operator vanishes for an on-shell photon and depends only logarithmically on the quantity $|k^2|/m^2$ for an off-shell photon with $k^2\neq 0$ \cite{Landau_b_4_1982}. However, other radiative corrections in vacuum like those corresponding to the vertex corrections show a double-logarithmic dependence on $|k^2|/m^2$ \cite{Landau_b_4_1982}. More closely to our result, the amplitudes of photon-photon scattering show a double-logarithmic dependence on the Mandelstam variable $s$ \cite{Landau_b_4_1982}, which corresponds to $2\theta_0 m^2$ in our notation [see also the remark below Eq. (\ref{P_b})]. On the other hand, we confirm that this logarithmic dependence on the energy scale is qualitatively different from the power-law dependence obtained in the CCF case in the limit $\kappa_0\to\infty$. Since for sufficiently large values of $\theta_0$, the parameter $r_0=\xi_0^2/\theta_0$ introduced above will be at a certain point smaller than unity, we can conclude that the polarization operator in a plane wave features a logarithmic behavior in the high-energy limit. 

As a byproduct of the above analysis, we can determine the expression of the total probability $P_{BW}(k)$ of nonlinear Breit-Wheeler pair production \cite{Reiss_1962,Nikishov_1964,Narozhny_2000,Roshchupkin_2001,Reiss_2009,Heinzl_2010b,Mueller_2011b,Titov_2012,Nousch_2012,Krajewska_2013b,Jansen_2013,Augustin_2014,Meuren_2015,Meuren_2016,Blackburn_2018c,Di_Piazza_2019d} in the same high-energy limit and for an unpolarized incoming photon, by applying the optical theorem \cite{Landau_b_4_1982,Meuren_2013}:
\begin{equation}
\label{P_BW_one_cycle}
P_{BW}=\frac{1}{m^2\theta_0}\text{Im}\left[\frac{P_e(k)+P_b(k)}{2}\right]=\frac{\alpha}{3}\frac{\xi_0^2}{\theta_0}\left[\log(\theta_0)-\gamma-1-\frac{3}{2}C_{Z,1}\right].
\end{equation}
We have explicitly verified that the same expression can be obtained starting from the probability of nonlinear Breit-Wheeler pair production as given, e.g., in Ref. \cite{Di_Piazza_2019d}. It is interesting to note that the dominating double logarithm appears only in the real part of the polarization operator. Instead, in the CCF limit one finds that both the real and the imaginary part of the polarization operator scale as $\kappa_0^{2/3}$ in the limit $\kappa_0\to\infty$.

Finally, we note that for any foreseeable laser intensity, in this limit the probability $P_{BW}$ is much smaller than unity as it is proportional to the small parameter $\alpha r_0\ll 1$ (assuming, of course, that at the energies under considerations $\alpha$ is much smaller than unity and that the logarithm does not compensate for the smallness of the quantity $\alpha r_0$).
%
%
\section{High-energy asymptotic of the one-loop mass operator in a plane wave}
The analysis of the asymptotic behavior of the leading-order mass operator in a plane wave (see Fig. 2) proceeds analogously as for the polarization operator. It is only technically more involved.

The starting point is the general expression of the leading-order mass operator found in Ref. \cite{Baier_1976_a}. However, analogously as in the previous section, we directly consider the ``diagonal'' part of the mass operator for incoming and outgoing electrons having the same on-shell four-momentum $p_1^{\mu}=p_2^{\mu}=p^{\mu}=(\varepsilon,\bm{p})$ ($p^2=m^2$) and average spin $\bm{\zeta}_1/2=\bm{\zeta}_2/2=\bm{\zeta}/2$ in their (common) rest frame. Analogously to the polarization operator, the vacuum part of the mass operator vanishes after renormalization \cite{Landau_b_4_1982}, whereas the field-dependent part can be written as $M_{f,\zeta}(p_1,p_2)=(2\pi)^3\delta^2(\bm{p}_{1,\perp}-\bm{p}_{2,\perp})\delta((k_0p_1)-(k_0p_2))M_{\zeta}(p)$, with $M_{\zeta}(p)=\sum_{j=1}^5M_{j,\zeta}(p)$. The five functions $M_{j,\zeta}(p)$ have the form \cite{Baier_1976_a}
\begin{align}
M_{1,\zeta}(p)&=\frac{\alpha}{2\pi}m\int_{-\infty}^{\infty}d\varphi\int_0^{\infty}\frac{d\tau}{\tau}\int_0^{\infty}\frac{dx}{(1+x)^2}\frac{1+2x}{1+x}\left\{e^{-i\frac{\tau x}{2\eta_0}[1+\xi_0^2\tilde{Q}^2(\varphi,\tau)]}-e^{-i\frac{\tau x}{2\eta_0}}\right\},\\
M_{2,\zeta}(p)&=\frac{\alpha}{4\pi}m\xi_0^2\int_{-\infty}^{\infty}d\varphi\int_0^{\infty}\frac{d\tau}{\tau}\int_0^{\infty}\frac{dx}{(1+x)^2}\Delta^2(\varphi,\tau)e^{-i\frac{\tau x}{2\eta_0}[1+\xi_0^2\tilde{Q}^2(\varphi,\tau)]},\\
M_{3,\zeta}(p)&=\frac{\alpha}{4\pi}m\xi_0^2\int_{-\infty}^{\infty}d\varphi\int_0^{\infty}\frac{d\tau}{\tau}\int_0^{\infty}\frac{dx}{(1+x)^2}\frac{x^2}{1+x}R(\varphi,\tau)e^{-i\frac{\tau x}{2\eta_0}[1+\xi_0^2\tilde{Q}^2(\varphi,\tau)]},\\
M_{4,\zeta}(p)&=\frac{\alpha}{4\pi}m\xi_0^2\int_{-\infty}^{\infty}d\varphi\int_0^{\infty}\frac{d\tau}{\tau}\int_0^{\infty}\frac{dx}{(1+x)^2}xS(\varphi,\tau)e^{-i\frac{\tau x}{2\eta_0}[1+\xi_0^2\tilde{Q}^2(\varphi,\tau)]},\\
M_{5,\zeta}(p)&=i\frac{\alpha}{4\pi}m\frac{(s_{\mu}f_0^{*\,\mu\nu}p_{\nu})}{m\eta_0}\int_{-\infty}^{\infty}d\varphi\int_0^{\infty}\frac{d\tau}{\tau}\int_0^{\infty}\frac{dx}{(1+x)^2}\frac{x}{1+x}\Delta(\varphi,\tau)e^{-i\frac{\tau x}{2\eta_0}[1+\xi_0^2\tilde{Q}^2(\varphi,\tau)]}.
\end{align}
Here, we have introduced the functions
\begin{align}
\label{Delta}
\Delta(\varphi,\tau)&=\psi(\varphi-\tau)-\psi(\varphi),\\
\label{tQ^2}
\tilde{Q}^2(\varphi,\tau)&=\frac{1}{\tau}\int_0^{\tau}d\tau'\Delta^2(\varphi,\tau')-\frac{1}{\tau^2}\left[\int_0^{\tau}d\tau'\Delta(\varphi,\tau')\right]^2,\\
\label{R}
R(\varphi,\tau)&=\left[\Delta(\varphi,\tau)-\frac{2}{\tau}\int_0^{\tau}d\tau'\Delta(\varphi,\tau')\right]\frac{1}{\tau}\int_0^{\tau}d\tau'\Delta(\varphi,\tau'),\\
\label{S}
S(\varphi,\tau)&=\frac{1}{\tau}\int_0^{\tau}d\tau'\Delta^2(\varphi,\tau'),
\end{align}
with
\begin{equation}
s^{\mu}=\left(\frac{\bm{p}\cdot\bm{\zeta}}{m},\bm{\zeta}+\frac{(\bm{p}\cdot\bm{\zeta})\bm{p}}{m(\varepsilon+m)}\right)
\end{equation}
being the spin four-vector \cite{Landau_b_4_1982} and $f_0^{*\,\mu\nu}=(1/2)\epsilon^{\mu\nu\lambda\rho}F_{0,\lambda\rho}/F_{cr}$, being the field pseudo-tensor amplitude in units of the critical field, and the parameter $\eta_0=(k_0p)/m^2$, which plays the same role as the parameter $\theta_0=(k_0k)/m^2$ in the case of the polarization operator. Note that only the term $M_{5,\zeta}(p)$ depends on the orientation of the average spin of the electron.

The strategy to find the high-energy asymptotic for $\eta_0\to \infty$ at $\xi_0$ constant is analogous to the one employed in the previous section. The integrals in $x$ have all the form
\begin{equation}
I_{n,d}(a)=\int_0^{\infty}\frac{dx}{(1+x)^2}\frac{x^n}{(1+x)^d}e^{-iax},
\end{equation}
with $n$ and $d$ being two non-negative integers and with $\text{Im}[a]<0$. By introducing the incomplete gamma function $\Gamma(0,z)$ \cite{NIST_b_2010}, the integrals that we need are
\begin{align}
I_{0,0}(a)&=1-iae^{ia}\Gamma(0,ia),\\
I_{0,1}(a)&=\frac{1}{2}[1-ia-a^2e^{ia}\Gamma(0,ia)],\\
I_{1,0}(a)&=-1+(1+ia)e^{ia}\Gamma(0,ia),\\
I_{1,1}(a)&=\frac{1}{2}[1+ia+a(a-2i)e^{ia}\Gamma(0,ia)],\\
I_{2,1}(a)&=\frac{1}{2}[-3-ia+(2+4ia-a^2)e^{ia}\Gamma(0,ia)].
\end{align}
In order to analyze the high-energy asymptotic behavior of each contribution $M_{j,\zeta}(p)$ to the mass operator, it is convenient now to introduce the quantities $\tilde{a}_0(\tau)=\tau/2\eta_0$ and $\tilde{a}_f(\varphi,\tau)=\tau\xi_0^2\tilde{Q}^2(\varphi,\tau)/2\eta_0$. As before, the strategy is based on the observation that the function $\tilde{a}_f(\varphi,\tau)$ for a finite pulse is bound, such that it vanishes in the high-energy limit $\eta_0\to\infty$ and $\xi_0$ fixed. Analogously to the case of the polarization operator, we first consider the leading-order contribution and we set $\tilde{a}_f(\varphi,\tau)=0$ in the terms from $M_{2,\zeta}(p)$ to $M_{5,\zeta}(p)$, whereas we approximate $I_{0,1}(\tilde{a}_0(\tau)+\tilde{a}_f(\varphi,\tau))-I_{0,1}(\tilde{a}_0(\tau))\approx I'_{0,1}(\tilde{a}_0(\tau))\tilde{a}_f(\varphi,\tau)$ and $I_{1,1}(\tilde{a}_0(\tau)+\tilde{a}_f(\varphi,\tau))-I_{1,1}(\tilde{a}_0(\tau))\approx I'_{1,1}(\tilde{a}_0(\tau))\tilde{a}_f(\varphi,\tau)$ in $M_{1,\zeta}(p)$. At this point, we perform the integrals in $\varphi$ and we already notice that the term $M_{5,\zeta}(p)$ vanishes. Now, we have shown that higher-order terms in $M_{5,\zeta}(p)$ in the expansion with respect to $\tilde{a}_f(\varphi,\tau)$ provide contributions subleading in $\eta_0$, such that we will ignore $M_{5,\zeta}(p)$ from now on. Concerning the other terms, we need the integrals
\begin{align}
\label{I_R}
\mathcal{I}_R(\tau)&=\int_{-\infty}^{\infty}d\varphi\,R(\varphi,\tau)=-\frac{2}{3}-\frac{8}{\tau^2}+\frac{4}{\sinh(\tau)}\left[\frac{\tau}{\sinh^2(\tau)}+\coth(\tau)+\frac{\tau}{2}\right],\\
\mathcal{I}_S(\tau)&=\int_{-\infty}^{\infty}d\varphi\,S(\varphi,\tau)=\frac{4}{3}+\frac{4}{\tau}\frac{1-\tau\coth(\tau)}{\sinh(\tau)},\\
\mathcal{I}_{\Delta^2}(\tau)&=\int_{-\infty}^{\infty}d\varphi\,\Delta^2(\varphi,\tau)=\frac{4}{3}+\frac{8}{\sinh(\tau)}\left[\frac{\tau}{\sinh^2(\tau)}-\coth(\tau)+\frac{\tau}{2}\right],\\
\label{I_tQ^2}
\mathcal{I}_{\tilde{Q}^2}(\tau)&=\int_{-\infty}^{\infty}d\varphi\,\tilde{Q}^2(\varphi,\tau)=\frac{2}{3}-\frac{4}{\tau^2}+\frac{4}{\tau}\frac{1}{\sinh(\tau)}.
\end{align}

Now, we start from the term $M_{1,\zeta}(p)$ and we can write it as
\begin{equation}
M_{1,\zeta}(p)=\frac{\alpha}{2\pi}m\xi_0^2\int_0^{\infty}d\tau\frac{d}{d\tau}\left[I_{0,1}\left(\frac{\tau}{2\eta_0}\right)+2I_{1,1}\left(\frac{\tau}{2\eta_0}\right)\right]\mathcal{I}_{\tilde{Q}^2}(\tau).
\end{equation}
From the asymptotic behavior of the integrand, we obtain that in the limit of large $\eta_0$ it is $M_{1,\zeta}(p)\approx -\alpha m\xi_0^2/2\pi$. 

We pass now to the term $M_{2,\zeta}(p)$, which is approximately given by
\begin{equation}
M_{2,\zeta}(p)=\frac{\alpha}{4\pi}m\xi_0^2\int_0^{\infty}\frac{d\tau}{\tau}\,I_{0,0}\left(\frac{\tau}{2\eta_0}\right)\mathcal{I}_{\Delta^2}(\tau).
\end{equation}
In this case it is necessary to split the integral by choosing a $\tau_0$ such that $1\ll\tau_0\ll\eta_0$. After approximating the functions of $\tau/\eta_0$ in the region $0\le\tau\le\tau_0$ for small values of the argument and the functions of $\tau$ in the region $\tau\ge\tau_0$ for large values of the argument, the final result is
\begin{equation}
M_{2,\zeta}(p)\approx\frac{\alpha}{3\pi}m\xi_0^2\left[\log(2\eta_0)-\gamma-i\frac{\pi}{2}-\frac{3}{4}C_{\Delta^2}\right],
\end{equation}
with
\begin{equation}
C_{\Delta^2}=\int_0^{\infty}d\tau\log(\tau)\mathcal{I}'_{\Delta^2}(\tau)\approx -0.637\ldots
\end{equation}

The asymptotic expressions of the terms
\begin{equation}
M_{3,\zeta}(p)=\frac{\alpha}{4\pi}m\xi_0^2\int_0^{\infty}\frac{d\tau}{\tau}\,I_{2,1}\left(\frac{\tau}{2\eta_0}\right)\mathcal{I}_R(\tau)
\end{equation}
and 
\begin{equation}
M_{4,\zeta}(p)=\frac{\alpha}{4\pi}m\xi_0^2\int_0^{\infty}\frac{d\tau}{\tau}\,I_{1,0}\left(\frac{\tau}{2\eta_0}\right)\mathcal{I}_S(\tau)
\end{equation}
can be determined with the same technique of splitting the integration region. We directly provide the asymptotic expressions:
\begin{align}
\begin{split}
M_{3,\zeta}(p)=&-\frac{\alpha}{12\pi}m\xi_0^2\bigg[\log^2(2\eta_0)-\left(3+2\gamma+i\pi-3C_{R,1}\right)\log(2\eta_0)\\
&\left.+1+3\gamma+\gamma^2+\frac{3}{2}i\pi+i\pi\gamma+\frac{\pi^2}{4}-\frac{3}{2}(3+2\gamma+i\pi)C_{R,1}-\frac{3}{2}C_{R,2}\right],
\end{split}\\
\begin{split}
M_{4,\zeta}(p)=&\frac{\alpha}{6\pi}m\xi_0^2\bigg[\log^2(2\eta_0)-\left(2+2\gamma+i\pi+\frac{3}{2}C_{S,1}\right)\log(2\eta_0)\\
&\left.+2\gamma+\gamma^2+i\pi+i\pi\gamma+\frac{\pi^2}{4}+\frac{3}{2}\left(1+\gamma+i\frac{\pi}{2}\right)C_{S,1}+\frac{3}{4}C_{S,2}\right],
\end{split}
\end{align}
where
\begin{align}
C_{R,1}&=\int_0^{\infty}d\tau\log(\tau)\mathcal{I}_R'(\tau)\approx -0.347\ldots,\\
C_{R,2}&=\int_0^{\infty}d\tau\log^2(\tau)\mathcal{I}_R'(\tau)\approx 0.154\ldots,\\
C_{S,1}&=\int_0^{\infty}d\tau\log(\tau)\mathcal{I}_S'(\tau)\approx 0.695\ldots,\\
C_{S,2}&=\int_0^{\infty}d\tau\log^2(\tau)\mathcal{I}_S'(\tau)\approx 1.02\ldots.
\end{align}
If we now analyze possible contributions of higher-order terms in the expansion with respect to $\tilde{a}_f(\varphi,\tau)$, it is easily recognized that only the terms $M_{3,\zeta}(p)$ and $M_{4,\zeta}(p)$ undergo corrections, which have to be taken into account here for consistency and which can be written as $\delta M_{3,\zeta}(p)=\alpha m\xi_0^2\tilde{\mathscr{I}}_R(\xi_0)/4\pi$ and $\delta M_{4,\zeta}(p)=\alpha m\xi_0^2\tilde{\mathscr{I}}_S(\xi_0)/4\pi$, with $\tilde{\mathscr{I}}_R(\xi_0)$ and $\tilde{\mathscr{I}}_S(\xi_0)$ depending only on the parameter $\xi_0$ (and on the pulse shape)
\begin{align}
\tilde{\mathscr{I}}_R(\xi_0)=&-\int_{-\infty}^{\infty}d\varphi \int_0^{\infty}\frac{d\tau}{\tau}R(\varphi,\tau)\log[1+\xi_0^2\tilde{Q}^2(\varphi,\tau)],\\
\tilde{\mathscr{I}}_S(\xi_0)=&-\int_{-\infty}^{\infty}d\varphi \int_0^{\infty}\frac{d\tau}{\tau}S(\varphi,\tau)\log[1+\xi_0^2\tilde{Q}^2(\varphi,\tau)].
\end{align}
In this way we obtain the complete asymptotics of the quantity $M_{\zeta}(p)=\sum_{j=1}^5M_{j,\zeta}(p)$ in the form
\begin{equation}
\begin{split}
M_{\zeta}(p)=&\frac{\alpha}{12\pi}m\xi_0^2\bigg\{\log^2(2\eta_0)+\left[3-2\gamma-i\pi-3(C_{R,1}+C_{S,1})\right]\log(2\eta_0)+3[\tilde{\mathscr{I}}_R(\xi_0)+\tilde{\mathscr{I}}_S(\xi_0)]\\
&+\frac{3}{2}[(2\gamma+i\pi)(C_{R,1}+C_{S,1})+3C_{R,1}+2C_{S,1}-2C_{\Delta^2}+C_{R,2}+C_{S,2}]\\
&\left.-7-3\gamma+\gamma^2-\frac{3}{2}i\pi+i\pi\gamma+\frac{\pi^2}{4}\right\}.
\end{split}
\end{equation}
The results above cannot be compared with the corresponding analytical asymptotic found in Ref. \cite{Baier_1976_a} in the case of a circularly polarized monochromatic field. However, by applying the same technique employed above, we have reproduced the asymptotic expression in Eq. (3.42) in Ref. \cite{Baier_1976_a}. It is interesting to observe that, due to the infinite extension of a monochromatic field, the asymptotic behavior of the mass operator is different from that found above. Although, in fact, the asymptotics shows a double-logarithmic behavior as here, the double-logarithm and the logarithm in the monochromatic case are evaluated at the effective parameter $\tilde{\eta}_0=(k_0p)/\tilde{m}^2$, where $\tilde{m}^2=m^2(1+\xi_0^2)$ is the effective electron mass in the circularly polarized laser field.

As in the case of the polarization operator, the asymptotic behavior of the mass operator in the high-energy limit $\theta_0\to\infty$ and $\xi_0$ fixed is logarithmic and qualitatively different from the power-law behavior in the CCF limit $\eta_0\to 0$ and $\xi_0\to\infty$ (such that $\chi_0=\eta_0\xi_0$ is finite) at large values of $\chi_0$ \cite{Ritus_1972}. In this case, the discriminating parameter between the two asymptotic behaviors is $s_0=\xi_0^2/\eta_0$, such that the high-energy limit requires that $s_0\ll 1$ whereas the CCF limit requires that $s_0\gg 1$. We recall that in the vacuum case nonzero radiative corrections in the mass operator (after renormalization) arise only for incoming electrons with an off-shell four-momentum $p^2\neq m^2$ and also increase logarithmically with the parameter $|p^2|/m^2$ \cite{Landau_b_4_1982}.

Analogously to the total probability of nonlinear Breit-Wheeler pair production in the case of the polarization operator, the imaginary part of the mass operator is related via the optical theorem to the total probability of nonlinear Compton scattering \cite{Ivanov_2004,Boca_2009,Harvey_2009,Mackenroth_2010,Boca_2011,Mackenroth_2011,Seipt_2011,Seipt_2011b,Dinu_2012,Krajewska_2012,Dinu_2013,Seipt_2013,Krajewska_2014,Wistisen_2014,Harvey_2015,Seipt_2016,Seipt_2016b,Angioi_2016,Harvey_2016b,Di_Piazza_2018c,Ilderton_2018b,Blackburn_2018}. In the case of an unpolarized incoming electron the high-energy asymptotic of the total probability $P_C$ reads
\begin{equation}
P_C=-\frac{2}{m\eta_0}\text{Im}M_0(p)=\frac{\alpha}{6}\frac{\xi_0^2}{\eta_0}\left[\log(2\eta_0)+\frac{3}{2}-\gamma-\frac{3}{2}(C_{R,1}+C_{S,1})\right].
\end{equation}

Finally, we also note here that for any foreseeable laser intensity, in this limit the probability $P_C$ is much smaller than unity as it is proportional to the small parameter $\alpha s_0\ll 1$ (we also assume here that at the energies under considerations $\alpha$ is much smaller than unity and that the logarithm does not compensate for the smallness of the quantity $\alpha s_0$).
%
%
\section{Generalization to arbitrary, finite pulse shapes}

In this section we generalize the above asymptotics to the case of an arbitrary pulse shape $\psi(\varphi)$ of the laser field, provided that it describes a finite pulse, i.e., that the integrals
\begin{align}
W_{\psi}&=\int_{-\infty}^{\infty}d\varphi\,\psi^2(\varphi),\\
W_{\psi'}&=\int_{-\infty}^{\infty}d\varphi\,\psi^{\prime\,2}(\varphi),\\
W_{\psi''}&=\int_{-\infty}^{\infty}d\varphi\,\psi^{\prime\prime\,2}(\varphi)
\end{align}
are finite and that
\begin{equation}
\lim_{\tau\to\pm\infty}\int_{-\infty}^{\infty}d\varphi\,\psi(\varphi)\psi(\varphi+\tau)=0.
\end{equation}
This last assumption plays a role in order to ascertain the behavior at large values of $\tau$ of integrals with respect to $\varphi$ involving, e.g., the functions $X(\varphi,\tau)$ and $Z(\varphi,\tau)$ in the case of the polarization operator. Note that all above integrals diverge for a monochromatic wave, such that the analysis below is inapplicable to this case.
\subsection{Polarization operator}
Below the same notation as in Sec. II is employed for the integrals $\mathcal{I}_X(\tau)$, $\mathcal{I}_Z(\tau)$, and $\mathcal{I}_{Q^2}(\tau)$ as in Eqs. (\ref{I_X})-(\ref{I_Q^2}) but of course with the general expressions in Eqs. (\ref{X})-(\ref{Q^2}). 

It is easily proved that for an arbitrary finite pulse the functions $\mathcal{I}_X(\tau)$, $\mathcal{I}_Z(\tau)$, and $\mathcal{I}_{Q^2}(\tau)$ tend to zero quadratically in the limit $\tau\to 0$ and, in particular, that
\begin{align}
\mathcal{I}_X(\tau)&\approx -W_{\psi'}\tau^2 && \text{for $\tau\ll 1$},\\
\mathcal{I}_Z(\tau)&\approx 2W_{\psi'}\tau^2 && \text{for $\tau\ll 1$},\\
\label{I_Q^2_l_tau}
\mathcal{I}_{Q^2}(\tau)&\approx \frac{1}{3}W_{\psi'}\tau^2 && \text{for $\tau\ll 1$}.
\end{align}
In the complementary limit $\tau\to\infty$ we instead obtain
\begin{align}
\lim_{\tau\to\infty}\mathcal{I}_X(\tau)&=0,\\
\lim_{\tau\to\infty}\mathcal{I}_Z(\tau)&= W_{\psi},\\
\lim_{\tau\to\infty}\mathcal{I}_{Q^2}(\tau)&= W_{\psi}.
\end{align}
These results already allow to carry out the asymptotic expansions of the integrals $\mathscr{I}_X$, $\mathscr{I}_Z$, and $\mathscr{I}_{Q^2}$ in Eqs. (\ref{int_I_X})-(\ref{int_I_Q^2}) as above. Starting from the leading-order expansion of the functions $I_n(a_0(\tau)+a_f(\varphi,\tau))$ with respect to $a_f(\varphi,\tau)$, we obtain
\begin{widetext}
\begin{align}
\mathscr{I}_X=&2\int_0^{\infty}\frac{d\tau}{\tau}\mathcal{I}_X(\tau),\\
\begin{split}
\mathscr{I}_Z=&\frac{W_{\psi}}{2}\left[\log^2(\theta_0)-\left(2\gamma+i\pi+\frac{2}{W_{\psi}}C_{Z,1}\right)\log(\theta_0)+\gamma^2+i\pi\gamma-\frac{5}{12}\pi^2-\log^22\right]\\
&+\left(\gamma+i\frac{\pi}{2}\right)C_{Z,1}+\frac{1}{2}C_{Z,2}+W_{\psi}C_K,
\end{split}\\
\mathscr{I}_{Q^2}=&\frac{2W_{\psi}}{\sqrt{\pi}}\left[\log(\theta_0)-2-\gamma-i\frac{\pi}{2}\right]-\frac{2}{\sqrt{\pi}}C_{Q^2,1},
\end{align}
\end{widetext}
where the definitions of the constants $C_{Z,1}$, $C_{Z,2}$, $C_K$, and $C_{Q^2,1}$ are as in Eqs. (\ref{C_Z_1})-(\ref{C_Q^2}) except, of course, that the numerical values of $C_{Z,1}$, $C_{Z,2}$, and $C_{Q^2,1}$ here depend on the pulse shape. 

Concerning the higher-order expansions of the functions $I_n(a_0(\tau)+a_f(\varphi,\tau))$ with respect to $a_f(\varphi,\tau)$, as it is clear from the discussion below Eq. (\ref{I_Z_n}), the results will be the same as above because they only depend on the properties of the functions $I_n(a_0(\tau)+a_f(\varphi,\tau))$. One has only to keep in mind that if the pulse has a duration corresponding to a phase $\Phi\gg 1$, then the asymptotics will be valid if $\theta_0\gg \Phi$. The reason is that in this case the functions $\mathcal{I}_{Z,n}(\tau)$ in Eq. (\ref{tilde_I_Z}) are significantly different from zero for $\tau\lesssim \Phi$ [see Eqs. (\ref{I_Z_n}) and (\ref{Q^2})] such that the asymptotic expansion of the derivatives $d^nI_1(a)/da^n$ at $a=4\tau/\theta_0$ for small values of the argument is valid only for $\theta_0/\Phi\gg 1$. Under this additional assumption, we obtain
\begin{align}
\label{P_e_G}
\begin{split}
P_e(k)=&-\frac{\alpha m^2\xi_0^2W_{\psi}}{2\pi}\left\{\log^2(\theta_0)-\left(2\gamma+i\pi+2+\frac{2}{W_{\psi}}C_{Z,1}\right)\log(\theta_0)+\frac{2}{W_{\psi}}\tilde{\mathscr{I}}_Z(\xi_0)+\gamma^2\right.\\
&+i\pi\gamma-\frac{5}{12}\pi^2-\log^22+4+2\left(\gamma+i\frac{\pi}{2}+C_K\right)\\
&\left.+\frac{2}{W_{\psi}}\left[\left(\gamma+i\frac{\pi}{2}\right)C_{Z,1}+\frac{1}{2}C_{Z,2}+C_{Q^2,1}+\mathscr{I}_X\right]\right\},
\end{split}\\
\label{P_b_G}
\begin{split}
P_b(k)=&P_e(k)+\frac{\alpha m^2\xi_0^2}{\pi}\mathscr{I}_X,
\end{split}
\end{align}
and
\begin{equation}
\label{P_BW}
P_{BW}=\frac{\alpha}{2}\frac{\xi_0^2W_{\psi}}{\theta_0}\left[\log(\theta_0)-\gamma-1-\frac{1}{W_{\psi}}C_{Z,1}\right].
\end{equation}
As a check about these results we recall that the angular frequency $\omega_0$ was introduced by hand in the expression of the polarization operator $P_f^{\mu\nu}(k_1,k_2)$ in Eq. (\ref{PO^munu}). Thus, this quantity must be effectively independent of $\omega_0$, which is the case if the coefficients $P_l(k)$ are proportional to $\omega_0$. This can indeed be verified by noticing in particular that $\xi_0^2\psi^2(\varphi)=e^2A_0^2(\varphi)/m^2$, that $\varphi=\omega_0(t-\bm{n}\cdot\bm{x})$ and that all the occurrences of $\omega_0$ in the logarithms of $\theta_0$ can be removed by means of a change of variable in the integrals in $\tau$ in the constants $C_{Z,1}$, $C_{Z,2}$, and $C_{Q^2,1}$ [see also Eqs. (\ref{C_Z_1})-(\ref{C_Q^2})].

As an additional test on the validity of our method, we show in Figs. 3 and
4 the real and the imaginary parts of the quantities $P_e(k)$ and $P_b(k)$ as
functions of $\theta_0$ evaluated according to the analytical asymptotics
in Eqs. (\ref{P_e_G}) and (\ref{P_b_G}), respectively, and to the exact expression in
Eqs. (\ref{P_e}) and (\ref{P_b}), respectively. The pulse-shape function $\psi(\varphi)=\sin^2(\varphi/2N)\sin(\varphi)$ for $\varphi\in [0,2N\pi]$ and zero otherwise has been employed. In order to show also the dependence of the results
on the pulse length, the results of two simulations are reported corresponding
to $N=5$ and $N=10$. Also, the parameter $\xi_0$ is assumed to be sufficiently small  as compared to unity that nonlinear contributions in $\xi_0^2$ to $P_e(k)$ and $P_b(k)$ can be neglected, and both the approximated and the exact expressions of $P_e(k)$ and $P_b(k)$ are effectively proportional to $\xi_0^2$. Thus, by conveniently plotting the quantities $P_e(k)$ and $P_b(k)$ in units of $-\alpha m^2\xi_0^2/\pi$, it is not necessary to specify a numerical value of $\xi_0$ (keeping in mind the assumption $\xi_0\ll 1$). The figures indicate the very good agreement between
the analytical asymptotics and the exact curves for large values of
$\theta_0$ and, as expected, an approximated linear dependence on $N$ [see Eqs. (\ref{P_e_G}) and (\ref{P_b_G})].
\begin{figure}
\begin{center}
\includegraphics[width=\columnwidth]{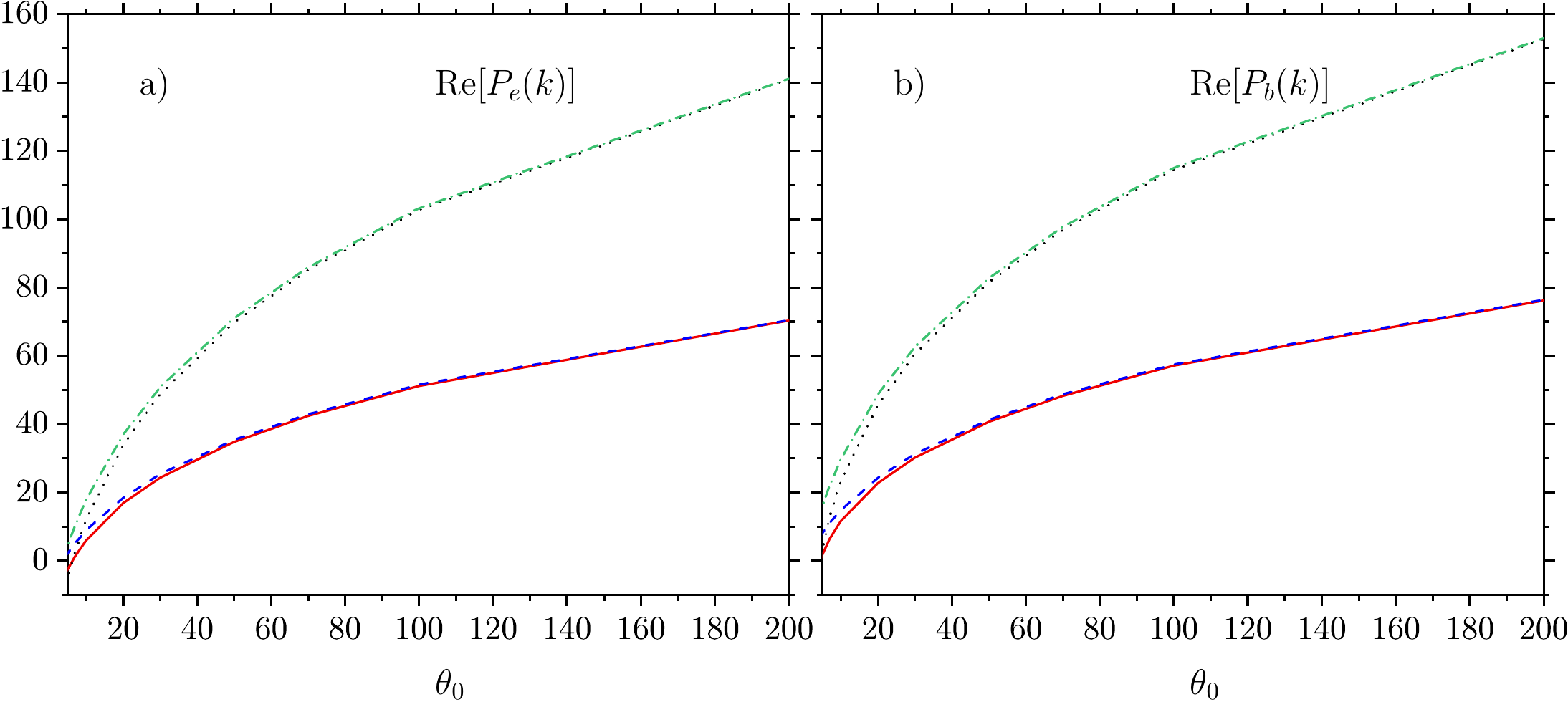}
\caption{Real part of the quantity $P_e(k)$ (Fig. 3a) and of the quantity $P_b(k)$ (Fig. 3b) in units of $-\alpha m^2\xi_0^2/\pi$. The pulse-shape function $\psi(\varphi)=\sin^2(\varphi/2N)\sin(\varphi)$ for $\varphi\in [0,2N\pi]$ and zero otherwise has been employed. In Fig. 3a the continuous red curve (dotted black curve) and the dashed blue curve (dash-dotted green curve) are obtained from the exact  expression in Eq. (\ref{P_e}) and from the asymptotic expression in Eq. (\ref{P_e_G}), respectively, and correspond to $N=5$ ($N=10$). The same colors and styles have been used for the curves in Fig. 3b with the exact and asymptotic expressions being given in Eq. (\ref{P_b}) and in Eq. (\ref{P_b_G}), respectively. The parameter $\xi_0$ is assumed to be sufficiently small that both the exact and the asymptotic expressions can be approximated to be proportional to $\xi_0^2$.}
\end{center}
\end{figure}
\begin{figure}
\begin{center}
\includegraphics[width=\columnwidth]{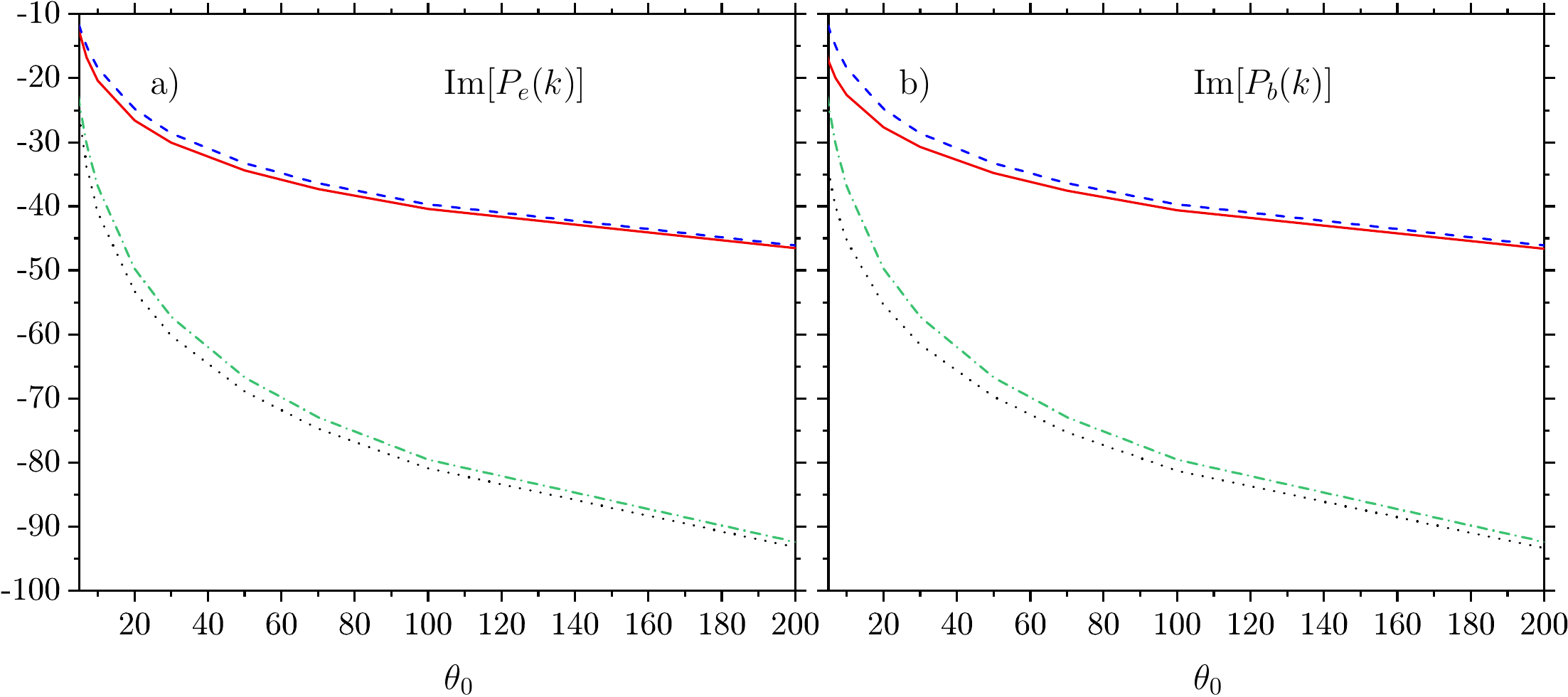}
\caption{Imaginary part of the quantity $P_e(k)$ (Fig. 4a) and of the quantity $P_b(k)$ (Fig. 4b) in units of $-\alpha m^2\xi_0^2/\pi$. The pulse-shape function $\psi(\varphi)=\sin^2(\varphi/2N)\sin(\varphi)$ for $\varphi\in [0,2N\pi]$ and zero otherwise has been employed. In Fig. 4a the continuous red curve (dotted black curve) and the dashed blue curve (dash-dotted green curve) are obtained from the exact  expression in Eq. (\ref{P_e}) and from the asymptotic expression in Eq. (\ref{P_e_G}), respectively, and correspond to $N=5$ ($N=10$). The same colors and styles have been used for the curves in Fig. 4b with the exact and asymptotic expressions being given in Eq. (\ref{P_b}) and in Eq. (\ref{P_b_G}), respectively. The parameter $\xi_0$ is assumed to be sufficiently small that both the exact and the asymptotic expressions can be approximated to be proportional to $\xi_0^2$.}
\end{center}
\end{figure}

Now, we observe that, since the probability $P_{BW}$ is proportional to $\xi_0^2$, one can ask whether it can be obtained starting from the cross section of linear Breit-Wheeler pair production \cite{Landau_b_4_1982} (note that the first two terms of the expansion of the probability of nonlinear Breit-Wheeler pair production for small $\xi_0$ and in a monochromatic plane wave can be found in Ref. \cite{Ritus_1985}). However, the result in Eq. (\ref{P_BW}) has been obtained under the assumption of a finite laser pulse, whereas the linear result is obtained for monochromatic photons. Thus, in order to reproduce Eq. (\ref{P_BW}) starting from the cross section of linear Breit-Wheeler pair production, one has to consider the incoming photon being in a coherent state according to the precise shape of the laser field. Conversely, one can start from the general results in Eqs. (\ref{P_e}) and (\ref{P_b}), expand the coefficients $P_e(k)$ and $P_b(k)$ for small $\xi_0$ up to terms of the order of $\xi_0^2$, and then employ a constant-amplitude pulse form $\psi(\varphi)=\cos(\varphi)$ for $-\Phi/2\le\varphi\le\Phi/2$ and zero otherwise, with $\Phi\gg 1$ (and ultimately sent to infinity when the monochromatic limit is considered). One finds that
\begin{align}
\label{I_X_m}
\mathcal{I}_X(\tau)&\sim-\left\{\frac{1}{2}\left[1-\frac{\sin^2(\tau)}{\tau^2}\right]+\frac{\sin(2\tau)}{2\tau}-\cos^2(\tau)\right\}\Phi,\\
\mathcal{I}_Z(\tau)&\sim\sin^2(\tau)\Phi,\\
\label{I_Q^2_m}
\mathcal{I}_{Q^2}(\tau)&\sim\frac{1}{2}\left[1-\frac{\sin^2(\tau)}{\tau^2}\right]\Phi,
\end{align}
where the symbol $\sim$ indicates that only the terms contributing to the diagonal part of the polarization operator in momentum space are retained. The asymptotic behavior of the functions $I_X(\tau)$ and $I_Z(\tau)$ for large $\tau$ is indeed very different from the corresponding ones in a finite pulse and this explains why one cannot obtain Eq. (\ref{P_BW}) from the cross section of linear Breit-Wheeler pair production. However, by employing the results in Eqs. (\ref{I_X_m})-(\ref{I_Q^2_m}), one finds that
\begin{align}
\begin{split}
P_e(k)=&-\frac{\alpha}{\pi}m^2\xi_0^2\Phi\int_0^{\infty}\frac{d\tau}{\tau}\int_1^{\infty}\frac{d\rho}{\rho^{3/2}}\frac{e^{-i\frac{4\tau\rho}{\theta_0}}}{\sqrt{\rho-1}}\\
&\qquad\times\left\langle\rho\sin^2(\tau)-\left\{\frac{3}{4}\left[1-\frac{\sin^2(\tau)}{\tau^2}\right]+\frac{\sin(2\tau)}{2\tau}-\cos^2(\tau)\right\}\right\rangle,
\end{split}\\
\begin{split}
P_b(k)=&-\frac{\alpha}{\pi}m^2\xi_0^2\Phi\int_0^{\infty}\frac{d\tau}{\tau}\int_1^{\infty}\frac{d\rho}{\rho^{3/2}}\frac{e^{-i\frac{4\tau\rho}{\theta_0}}}{\sqrt{\rho-1}}\left\{\rho\sin^2(\tau)-\frac{1}{4}\left[1-\frac{\sin^2(\tau)}{\tau^2}\right]\right\}.
\end{split}
\end{align}
The asymptotic expressions for large values of $\theta_0$ are obtained by working out first the integral in $\tau$ and then that in $\rho$ as above, and the result for the pair-production probability is
\begin{equation}
P_{BW}=\frac{\alpha}{4}\frac{\xi_0^2\Phi}{\theta_0}\left[\log(2\theta_0)-1\right].
\end{equation}
This expression can indeed be obtained starting from the cross section of linear Breit-Wheeler pair production as given in Ref. \cite{Landau_b_4_1982} and taking into account the flux of laser photons.

\subsection{Mass operator}

We can follow a similar reasoning in the case of the mass operator and of the functions $\mathcal{I}_R(\tau)$, $\mathcal{I}_S(\tau)$, $\mathcal{I}_{\Delta^2}(\tau)$, and $\mathcal{I}_{\tilde{Q}^2}(\tau)$ introduced in Eqs. (\ref{I_R})-(\ref{I_tQ^2}) and to be intended below according to the general definitions in Eqs. (\ref{Delta})-(\ref{S}). One can easily show that under the already mentioned conditions on the pulse function $\psi(\varphi)$, one obtains
\begin{align}
\mathcal{I}_R(\tau)&\approx -\frac{1}{72}W_{\psi''}\tau^4 && \text{for $\tau\ll 1$},\\
\mathcal{I}_S(\tau)&\approx \frac{1}{3}W_{\psi'}\tau^2 && \text{for $\tau\ll 1$},\\
\mathcal{I}_{\Delta^2}(\tau)&\approx W_{\psi'}\tau^2 && \text{for $\tau\ll 1$},\\
\mathcal{I}_{\tilde{Q}^2}(\tau)&\approx \frac{1}{12}W_{\psi'}\tau^2 && \text{for $\tau\ll 1$},
\end{align}
and
\begin{align}
\lim_{\tau\to\infty}\mathcal{I}_R(\tau)&=-W_{\psi},\\
\lim_{\tau\to\infty}\mathcal{I}_S(\tau)&=2W_{\psi},\\
\lim_{\tau\to\infty}\mathcal{I}_{\Delta^2}(\tau)&=2W_{\psi},\\
\lim_{\tau\to\infty}\mathcal{I}_{\tilde{Q}^2}(\tau)&=W_{\psi}.
\end{align}
Based on these results and on the results of Sec. III, it is straightforward to generalize the asymptotic expressions of the terms $M_{1,\zeta}(p)$, $\ldots$, $M_{4,\zeta}(p)$ up to the leading order in the expansion of the functions $I_{n,d}(\tilde{a}_0(\tau)+\tilde{a}_f(\varphi,\tau))$ for small values of $\tilde{a}_f(\varphi,\tau)$:
\begin{align}
M_{1,\zeta}(p)=&-\frac{3\alpha}{4\pi}m\xi_0^2W_{\psi},\\
M_{2,\zeta}(p)=&\frac{\alpha}{2\pi}m\xi_0^2\left\{W_{\psi}\left[\log(2\eta_0)-\gamma-i\frac{\pi}{2}\right]-\frac{1}{2}C_{\Delta^2}\right\},\\
\begin{split}
M_{3,\zeta}(p)=&-\frac{\alpha}{8\pi}m\xi_0^2W_{\psi}\bigg[\log^2(2\eta_0)-\left(3+2\gamma+i\pi-\frac{2}{W_{\psi}}C_{R,1}\right)\log(2\eta_0)\\
&\left.+1+3\gamma+\gamma^2+\frac{3}{2}i\pi+i\pi\gamma+\frac{\pi^2}{4}-\frac{1}{W_{\psi}}(3+2\gamma+i\pi)C_{R,1}-\frac{1}{W_{\psi}}C_{R,2}\right],
\end{split}\\
\begin{split}
M_{4,\zeta}(p)=&\frac{\alpha}{4\pi}m\xi_0^2W_{\psi}\bigg[\log^2(2\eta_0)-\left(2+2\gamma+i\pi+\frac{1}{W_{\psi}}C_{S,1}\right)\log(2\eta_0)\\
&\left.+2\gamma+\gamma^2+i\pi+i\pi\gamma+\frac{\pi^2}{4}+\frac{1}{W_{\psi}}\left(1+\gamma+i\frac{\pi}{2}\right)C_{S,1}+\frac{1}{2W_{\psi}}C_{S,2}\right].
\end{split}
\end{align}
In this way the final result for the function $M_{\zeta}(p)$ also including the contributions of high-order terms in $\tilde{a}_f(\varphi,\tau)$ reads
\begin{equation}
\label{M}
\begin{split}
M_{\zeta}(p)=&\frac{\alpha}{8\pi}m\xi_0^2W_{\psi}\bigg\{\log^2(2\eta_0)+\left[3-2\gamma-i\pi-\frac{2}{W_{\psi}}(C_{R,1}+C_{S,1})\right]\log(2\eta_0)\\
&+\frac{2}{W_{\psi}}[\tilde{\mathscr{I}}_R(\xi_0)+\tilde{\mathscr{I}}_S(\xi_0)]-7-3\gamma+\gamma^2-\frac{3}{2}i\pi+i\pi\gamma+\frac{\pi^2}{4}\\
&\left.+\frac{1}{W_{\psi}}[(2\gamma+i\pi)(C_{R,1}+C_{S,1})+3C_{R,1}+2C_{S,1}-2C_{\Delta^2}+C_{R,2}+C_{S,2}]\right\},
\end{split}
\end{equation}
with all constants being defined as in Sec. III but, of course, for a general pulse-shape function $\psi(\varphi)$. 

Finally, the asymptotic of the total probability of nonlinear Compton scattering in an arbitrary finite pulse at high-energies reads
\begin{equation}
\label{P_C}
P_C=\frac{\alpha}{4}\frac{\xi_0^2W_{\psi}}{\eta_0}\left[\log(2\eta_0)+\frac{3}{2}-\gamma-\frac{1}{W_{\psi}}(C_{R,1}+C_{S,1})\right].
\end{equation}
The two remarks about the appearance of the angular frequency $\omega_0$ and the agreement with the cross section of the corresponding linear process (in this case linear Compton scattering) can be verified also in Eqs. (\ref{M}) and (\ref{P_C}) (note that the first two terms of the expansion of the probability of nonlinear Compton scattering for small $\xi_0$ and in a monochromatic plane wave can be found in Ref. \cite{Ritus_1985}).

\subsection{An additional remark}
The results in Eqs. (\ref{P_e_G}), (\ref{P_b_G}), and (\ref{M}) also confirm that the high-energy behavior of the polarization (mass) operator depends logarithmically on the center-of-momentum energy of the incoming photon (electron) and a laser photon in a qualitatively similar way as in vacuum. This behavior is consequently very different from the power-law behavior observed in the CCF limit at large $\kappa_0$ ($\chi_0$). In this respect, we would like to point out also that, although the above theoretical analysis reconciles the high-energy behavior of QED in vacuum and of QED in an intense plane wave, it does not prevent the experimental verification of the interesting regime where the RN conjecture would apply \cite{Blackburn_2018b,Baumann_2018,Yakimenko_2018}. In fact, according to the above results, if the parametric conditions $\theta_0\ll 1$, $\xi_0\gg 1$, and $\kappa_0=\theta_0\xi_0\gg 1$ (in the case of an incoming photon) or $\eta_0\ll 1$, $\xi_0\gg 1$, and $\chi_0=\eta_0\xi_0\gg 1$ (in the case of an incoming electron) are fulfilled at the given experimental conditions, then the parameters $r_0=\xi_0^2/\theta_0=\kappa_0^2/\theta_0^3$ and $s_0=\xi_0^2/\eta_0=\chi_0^2/\eta_0^3$ are automatically much larger than unity and, according to the RN conjecture, the power-law increase of the effective coupling constant can in principle be tested.
\section{Conclusions and outlook}
In conclusion, we have shown that the one-loop polarization operator and mass operator in an intense, finite plane wave feature a logarithmic behavior at high energies, similar to other radiative corrections in vacuum. This is qualitatively different from the power-law behavior in the regions $\kappa_0\gg 1$ and $\chi_0\gg 1$, which is observed within the CCF limit. The difference arises from the non-commutativity of the high-energy limit (either $\theta_0\to \infty$ and $\xi_0$ fixed or $\eta_0\to \infty$ and $\xi_0$ fixed) and of the CCF limit (either $\theta_0\to 0$ at $\kappa_0$ fixed or $\eta_0\to 0$ at $\chi_0$ fixed). In the case of the polarization operator and of the mass operator the discriminating parameters between the two regimes have been identified to be $r_0=\xi_0^2/\theta_0$ and $s_0=\xi_0^2/\eta_0$, respectively, which should be much smaller (larger) than unity in order the high-energy (low-frequency/CCF) limit to apply. 

As a byproduct we have obtained the high-energy asymptotics of the total probabilities of nonlinear Breit-Wheeler pair production and of nonlinear Compton scattering for an unpolarized incoming photon and electron, respectively, and for an arbitrary, finite plane-wave field. Both these probabilities are proportional to $\xi_0^2$, and depend as $[A_{BW}\log(\theta_0)+B_{BW}]/\theta_0$ on $\theta_0$ (the pair-production probability) and as $[A_C\log(\eta_0)+B_C]/\eta_0$ on $\eta_0$ (the photon emission probability), with the values of the constants $A_{BW}$, $B_{BW}$, $A_C$, and $B_C$ depending on the pulse shape.

The above analysis was carried out by considering an on-shell incoming particle for a more consistent comparison with available results in a CCF also obtained for on-shell incoming particles. In the vacuum case the corresponding radiative corrections vanish after renormalization and the high-energy behavior of the radiative corrections refers to incoming particles with larger and larger ``virtualities'', parametrized by the quantity $|q^2|/m^2$, with $q^{\mu}$ being the corresponding off-shell four-momentum. The analysis of this different asymptotic region is extremely interesting but goes beyond the present study, and will be the subject of a future investigation.

%


\begin{thebibliography}{81}%
\makeatletter
\providecommand \@ifxundefined [1]{%
 \@ifx{#1\undefined}
}%
\providecommand \@ifnum [1]{%
 \ifnum #1\expandafter \@firstoftwo
 \else \expandafter \@secondoftwo
 \fi
}%
\providecommand \@ifx [1]{%
 \ifx #1\expandafter \@firstoftwo
 \else \expandafter \@secondoftwo
 \fi
}%
\providecommand \natexlab [1]{#1}%
\providecommand \enquote  [1]{``#1''}%
\providecommand \bibnamefont  [1]{#1}%
\providecommand \bibfnamefont [1]{#1}%
\providecommand \citenamefont [1]{#1}%
\providecommand \href@noop [0]{\@secondoftwo}%
\providecommand \href [0]{\begingroup \@sanitize@url \@href}%
\providecommand \@href[1]{\@@startlink{#1}\@@href}%
\providecommand \@@href[1]{\endgroup#1\@@endlink}%
\providecommand \@sanitize@url [0]{\catcode `\\12\catcode `\$12\catcode
  `\&12\catcode `\#12\catcode `\^12\catcode `\_12\catcode `\%12\relax}%
\providecommand \@@startlink[1]{}%
\providecommand \@@endlink[0]{}%
\providecommand \url  [0]{\begingroup\@sanitize@url \@url }%
\providecommand \@url [1]{\endgroup\@href {#1}{\urlprefix }}%
\providecommand \urlprefix  [0]{URL }%
\providecommand \Eprint [0]{\href }%
\providecommand \doibase [0]{http://dx.doi.org/}%
\providecommand \selectlanguage [0]{\@gobble}%
\providecommand \bibinfo  [0]{\@secondoftwo}%
\providecommand \bibfield  [0]{\@secondoftwo}%
\providecommand \translation [1]{[#1]}%
\providecommand \BibitemOpen [0]{}%
\providecommand \bibitemStop [0]{}%
\providecommand \bibitemNoStop [0]{.\EOS\space}%
\providecommand \EOS [0]{\spacefactor3000\relax}%
\providecommand \BibitemShut  [1]{\csname bibitem#1\endcsname}%
\let\auto@bib@innerbib\@empty
\bibitem [{\citenamefont {Hanneke}\ \emph {et~al.}(2008)\citenamefont
  {Hanneke}, \citenamefont {Fogwell},\ and\ \citenamefont
  {Gabrielse}}]{Hanneke_2008}%
  \BibitemOpen
  \bibfield  {author} {\bibinfo {author} {\bibfnamefont {D.}~\bibnamefont
  {Hanneke}}, \bibinfo {author} {\bibfnamefont {S.}~\bibnamefont {Fogwell}}, \
  and\ \bibinfo {author} {\bibfnamefont {G.}~\bibnamefont {Gabrielse}},\
  }\href@noop {} {\bibfield  {journal} {\bibinfo  {journal} {Phys. Rev. Lett.}\
  }\textbf {\bibinfo {volume} {100}},\ \bibinfo {pages} {120801} (\bibinfo
  {year} {2008})}\BibitemShut {NoStop}%
\bibitem [{\citenamefont {Sturm}\ \emph {et~al.}(2011)\citenamefont {Sturm},
  \citenamefont {Wagner}, \citenamefont {Schabinger}, \citenamefont {Zatorski},
  \citenamefont {Harman}, \citenamefont {Quint}, \citenamefont {Werth},
  \citenamefont {Keitel},\ and\ \citenamefont {Blaum}}]{Sturm_2011}%
  \BibitemOpen
  \bibfield  {author} {\bibinfo {author} {\bibfnamefont {S.}~\bibnamefont
  {Sturm}}, \bibinfo {author} {\bibfnamefont {A.}~\bibnamefont {Wagner}},
  \bibinfo {author} {\bibfnamefont {B.}~\bibnamefont {Schabinger}}, \bibinfo
  {author} {\bibfnamefont {J.}~\bibnamefont {Zatorski}}, \bibinfo {author}
  {\bibfnamefont {Z.}~\bibnamefont {Harman}}, \bibinfo {author} {\bibfnamefont
  {W.}~\bibnamefont {Quint}}, \bibinfo {author} {\bibfnamefont
  {G.}~\bibnamefont {Werth}}, \bibinfo {author} {\bibfnamefont {C.~H.}\
  \bibnamefont {Keitel}}, \ and\ \bibinfo {author} {\bibfnamefont
  {K.}~\bibnamefont {Blaum}},\ }\href@noop {} {\bibfield  {journal} {\bibinfo
  {journal} {Phys. Rev. Lett.}\ }\textbf {\bibinfo {volume} {107}},\ \bibinfo
  {pages} {023002} (\bibinfo {year} {2011})}\BibitemShut {NoStop}%
\bibitem [{\citenamefont {Jauch}\ and\ \citenamefont
  {Rohrlich}(1976)}]{Jauch_b_1976}%
  \BibitemOpen
  \bibfield  {author} {\bibinfo {author} {\bibfnamefont {J.~M.}\ \bibnamefont
  {Jauch}}\ and\ \bibinfo {author} {\bibfnamefont {F.}~\bibnamefont
  {Rohrlich}},\ }\href@noop {} {\emph {\bibinfo {title} {The Theory of Photons
  and Electrons}}}\ (\bibinfo  {publisher} {Springer, Berlin},\ \bibinfo {year}
  {1976})\BibitemShut {NoStop}%
\bibitem [{\citenamefont {Itzykson}\ and\ \citenamefont
  {Zuber}(1980)}]{Itzykson_b_1980}%
  \BibitemOpen
  \bibfield  {author} {\bibinfo {author} {\bibfnamefont {C.}~\bibnamefont
  {Itzykson}}\ and\ \bibinfo {author} {\bibfnamefont {J.-B.}\ \bibnamefont
  {Zuber}},\ }\href@noop {} {\emph {\bibinfo {title} {Quantum Field Theory}}}\
  (\bibinfo  {publisher} {McGraw-Hill Inc., New York},\ \bibinfo {year}
  {1980})\BibitemShut {NoStop}%
\bibitem [{\citenamefont {Berestetskii}\ \emph {et~al.}(1982)\citenamefont
  {Berestetskii}, \citenamefont {Lifshitz},\ and\ \citenamefont
  {Pitaevskii}}]{Landau_b_4_1982}%
  \BibitemOpen
  \bibfield  {author} {\bibinfo {author} {\bibfnamefont {V.~B.}\ \bibnamefont
  {Berestetskii}}, \bibinfo {author} {\bibfnamefont {E.~M.}\ \bibnamefont
  {Lifshitz}}, \ and\ \bibinfo {author} {\bibfnamefont {L.~P.}\ \bibnamefont
  {Pitaevskii}},\ }\href@noop {} {\emph {\bibinfo {title} {Quantum
  Electrodynamics}}}\ (\bibinfo  {publisher} {Elsevier Butterworth-Heinemann,
  Oxford},\ \bibinfo {year} {1982})\BibitemShut {NoStop}%
\bibitem [{\citenamefont {Schwartz}(2014)}]{Schwartz_b_2014}%
  \BibitemOpen
  \bibfield  {author} {\bibinfo {author} {\bibfnamefont {M.}~\bibnamefont
  {Schwartz}},\ }\href@noop {} {\emph {\bibinfo {title} {Quantum Field Theory
  and the Standard Model}}}\ (\bibinfo  {publisher} {Cambridge University
  Press, Cambridge},\ \bibinfo {year} {2014})\BibitemShut {NoStop}%
\bibitem [{\citenamefont {Fradkin}\ \emph {et~al.}(1991)\citenamefont
  {Fradkin}, \citenamefont {Gitman},\ and\ \citenamefont
  {Shvartsman}}]{Fradkin_b_1991}%
  \BibitemOpen
  \bibfield  {author} {\bibinfo {author} {\bibfnamefont {E.~S.}\ \bibnamefont
  {Fradkin}}, \bibinfo {author} {\bibfnamefont {D.~M.}\ \bibnamefont {Gitman}},
  \ and\ \bibinfo {author} {\bibfnamefont {{\relax Sh}.~M.}\ \bibnamefont
  {Shvartsman}},\ }\href@noop {} {\emph {\bibinfo {title} {Quantum
  Electrodynamics with Unstable Vacuum}}}\ (\bibinfo  {publisher} {Springer,
  Berlin},\ \bibinfo {year} {1991})\BibitemShut {NoStop}%
\bibitem [{\citenamefont {Dittrich}\ and\ \citenamefont
  {Reuter}(1985)}]{Dittrich_b_1985}%
  \BibitemOpen
  \bibfield  {author} {\bibinfo {author} {\bibfnamefont {W.}~\bibnamefont
  {Dittrich}}\ and\ \bibinfo {author} {\bibfnamefont {M.}~\bibnamefont
  {Reuter}},\ }\href@noop {} {\emph {\bibinfo {title} {Effective Lagrangians in
  Quantum Electrodynamics}}}\ (\bibinfo  {publisher} {Springer, Heidelberg},\
  \bibinfo {year} {1985})\BibitemShut {NoStop}%
\bibitem [{\citenamefont {Yanovsky}\ \emph {et~al.}(2008)\citenamefont
  {Yanovsky}, \citenamefont {Chvykov}, \citenamefont {Kalinchenko},
  \citenamefont {Rousseau}, \citenamefont {Planchon}, \citenamefont {Matsuoka},
  \citenamefont {Maksimchuk}, \citenamefont {Nees}, \citenamefont
  {Ch\'{e}riaux}, \citenamefont {Mourou},\ and\ \citenamefont
  {Krushelnick}}]{Yanovsky_2008}%
  \BibitemOpen
  \bibfield  {author} {\bibinfo {author} {\bibfnamefont {V.}~\bibnamefont
  {Yanovsky}}, \bibinfo {author} {\bibfnamefont {V.}~\bibnamefont {Chvykov}},
  \bibinfo {author} {\bibfnamefont {G.}~\bibnamefont {Kalinchenko}}, \bibinfo
  {author} {\bibfnamefont {P.}~\bibnamefont {Rousseau}}, \bibinfo {author}
  {\bibfnamefont {T.}~\bibnamefont {Planchon}}, \bibinfo {author}
  {\bibfnamefont {T.}~\bibnamefont {Matsuoka}}, \bibinfo {author}
  {\bibfnamefont {A.}~\bibnamefont {Maksimchuk}}, \bibinfo {author}
  {\bibfnamefont {J.}~\bibnamefont {Nees}}, \bibinfo {author} {\bibfnamefont
  {G.}~\bibnamefont {Ch\'{e}riaux}}, \bibinfo {author} {\bibfnamefont
  {G.}~\bibnamefont {Mourou}}, \ and\ \bibinfo {author} {\bibfnamefont
  {K.}~\bibnamefont {Krushelnick}},\ }\href@noop {} {\bibfield  {journal}
  {\bibinfo  {journal} {Opt. Express}\ }\textbf {\bibinfo {volume} {16}},\
  \bibinfo {pages} {2109} (\bibinfo {year} {2008})}\BibitemShut {NoStop}%
\bibitem [{\citenamefont {Papadopoulos}\ \emph {et~al.}(2016)\citenamefont
  {Papadopoulos}, \citenamefont {Zou}, \citenamefont {Le~Blanc}, \citenamefont
  {Ch\'{e}riaux}, \citenamefont {Georges}, \citenamefont {Druon}, \citenamefont
  {Mennerat}, \citenamefont {Ramirez}, \citenamefont {Martin}, \citenamefont
  {Fr\'{e}neaux}, \citenamefont {Beluze}, \citenamefont {Lebas}, \citenamefont
  {Monot}, \citenamefont {Mathieu},\ and\ \citenamefont
  {Audebert}}]{APOLLON_10P}%
  \BibitemOpen
  \bibfield  {author} {\bibinfo {author} {\bibfnamefont {D.}~\bibnamefont
  {Papadopoulos}}, \bibinfo {author} {\bibfnamefont {J.}~\bibnamefont {Zou}},
  \bibinfo {author} {\bibfnamefont {C.}~\bibnamefont {Le~Blanc}}, \bibinfo
  {author} {\bibfnamefont {G.}~\bibnamefont {Ch\'{e}riaux}}, \bibinfo {author}
  {\bibfnamefont {P.}~\bibnamefont {Georges}}, \bibinfo {author} {\bibfnamefont
  {F.}~\bibnamefont {Druon}}, \bibinfo {author} {\bibfnamefont
  {G.}~\bibnamefont {Mennerat}}, \bibinfo {author} {\bibfnamefont
  {P.}~\bibnamefont {Ramirez}}, \bibinfo {author} {\bibfnamefont
  {L.}~\bibnamefont {Martin}}, \bibinfo {author} {\bibfnamefont
  {A.}~\bibnamefont {Fr\'{e}neaux}}, \bibinfo {author} {\bibfnamefont
  {A.}~\bibnamefont {Beluze}}, \bibinfo {author} {\bibfnamefont
  {N.}~\bibnamefont {Lebas}}, \bibinfo {author} {\bibfnamefont
  {P.}~\bibnamefont {Monot}}, \bibinfo {author} {\bibfnamefont
  {F.}~\bibnamefont {Mathieu}}, \ and\ \bibinfo {author} {\bibfnamefont
  {P.}~\bibnamefont {Audebert}},\ }\href@noop {} {\bibfield  {journal}
  {\bibinfo  {journal} {High Power Laser Sci. Eng.}\ }\textbf {\bibinfo
  {volume} {4}},\ \bibinfo {pages} {e34} (\bibinfo {year} {2016})}\BibitemShut
  {NoStop}%
\bibitem [{\citenamefont {{Extreme Light Infrastructure (ELI),
  https://eli-laser.eu/}}(2017)}]{ELI}%
  \BibitemOpen
  \bibfield  {author} {\bibinfo {author} {\bibnamefont {{Extreme Light
  Infrastructure (ELI), https://eli-laser.eu/}}},\ }\href@noop {} {} (\bibinfo
  {year} {2017})\BibitemShut {NoStop}%
\bibitem [{\citenamefont {{Center for Relativistic Laser Science (CoReLS),
  https://www.ibs.re.kr/eng/sub02\_03\_05.do}}(2017)}]{CoReLS}%
  \BibitemOpen
  \bibfield  {author} {\bibinfo {author} {\bibnamefont {{Center for
  Relativistic Laser Science (CoReLS),
  https://www.ibs.re.kr/eng/sub02\_03\_05.do}}},\ }\href@noop {} {} (\bibinfo
  {year} {2017})\BibitemShut {NoStop}%
\bibitem [{\citenamefont {Mitter}(1975)}]{Mitter_1975}%
  \BibitemOpen
  \bibfield  {author} {\bibinfo {author} {\bibfnamefont {H.}~\bibnamefont
  {Mitter}},\ }\href@noop {} {\bibfield  {journal} {\bibinfo  {journal} {Acta
  Phys. Austriaca}\ }\textbf {\bibinfo {volume} {XIV}},\ \bibinfo {pages} {397}
  (\bibinfo {year} {1975})}\BibitemShut {NoStop}%
\bibitem [{\citenamefont {Ritus}(1985)}]{Ritus_1985}%
  \BibitemOpen
  \bibfield  {author} {\bibinfo {author} {\bibfnamefont {V.~I.}\ \bibnamefont
  {Ritus}},\ }\href@noop {} {\bibfield  {journal} {\bibinfo  {journal} {J. Sov.
  Laser Res.}\ }\textbf {\bibinfo {volume} {6}},\ \bibinfo {pages} {497}
  (\bibinfo {year} {1985})}\BibitemShut {NoStop}%
\bibitem [{\citenamefont {Ehlotzky}\ \emph {et~al.}(2009)\citenamefont
  {Ehlotzky}, \citenamefont {Krajewska},\ and\ \citenamefont
  {Kami\'{n}ski}}]{Ehlotzky_2009}%
  \BibitemOpen
  \bibfield  {author} {\bibinfo {author} {\bibfnamefont {F.}~\bibnamefont
  {Ehlotzky}}, \bibinfo {author} {\bibfnamefont {K.}~\bibnamefont {Krajewska}},
  \ and\ \bibinfo {author} {\bibfnamefont {J.~Z.}\ \bibnamefont
  {Kami\'{n}ski}},\ }\href@noop {} {\bibfield  {journal} {\bibinfo  {journal}
  {Rep. Prog. Phys.}\ }\textbf {\bibinfo {volume} {72}},\ \bibinfo {pages}
  {046401} (\bibinfo {year} {2009})}\BibitemShut {NoStop}%
\bibitem [{\citenamefont {Reiss}(2009)}]{Reiss_2009}%
  \BibitemOpen
  \bibfield  {author} {\bibinfo {author} {\bibfnamefont {H.~R.}\ \bibnamefont
  {Reiss}},\ }\href@noop {} {\bibfield  {journal} {\bibinfo  {journal} {Eur.
  Phys. J. D}\ }\textbf {\bibinfo {volume} {55}},\ \bibinfo {pages} {365}
  (\bibinfo {year} {2009})}\BibitemShut {NoStop}%
\bibitem [{\citenamefont {Di~Piazza}\ \emph {et~al.}(2012)\citenamefont
  {Di~Piazza}, \citenamefont {M\"{u}ller}, \citenamefont {Hatsagortsyan},\ and\
  \citenamefont {Keitel}}]{Di_Piazza_2012}%
  \BibitemOpen
  \bibfield  {author} {\bibinfo {author} {\bibfnamefont {A.}~\bibnamefont
  {Di~Piazza}}, \bibinfo {author} {\bibfnamefont {C.}~\bibnamefont
  {M\"{u}ller}}, \bibinfo {author} {\bibfnamefont {K.~Z.}\ \bibnamefont
  {Hatsagortsyan}}, \ and\ \bibinfo {author} {\bibfnamefont {C.~H.}\
  \bibnamefont {Keitel}},\ }\href@noop {} {\bibfield  {journal} {\bibinfo
  {journal} {Rev. Mod. Phys.}\ }\textbf {\bibinfo {volume} {84}},\ \bibinfo
  {pages} {1177} (\bibinfo {year} {2012})}\BibitemShut {NoStop}%
\bibitem [{\citenamefont {Dunne}(2014)}]{Dunne_2014}%
  \BibitemOpen
  \bibfield  {author} {\bibinfo {author} {\bibfnamefont {G.}~\bibnamefont
  {Dunne}},\ }\href@noop {} {\bibfield  {journal} {\bibinfo  {journal} {Eur.
  Phys. J. Special Topics}\ }\textbf {\bibinfo {volume} {223}},\ \bibinfo
  {pages} {1055} (\bibinfo {year} {2014})}\BibitemShut {NoStop}%
\bibitem [{\citenamefont {Bula}\ \emph {et~al.}(1996)\citenamefont {Bula},
  \citenamefont {McDonald}, \citenamefont {Prebys}, \citenamefont {Bamber},
  \citenamefont {Boege}, \citenamefont {Kotseroglou}, \citenamefont
  {Melissinos}, \citenamefont {Meyerhofer}, \citenamefont {Ragg}, \citenamefont
  {Burke}, \citenamefont {Field}, \citenamefont {Horton-Smith}, \citenamefont
  {Odian}, \citenamefont {Spencer}, \citenamefont {Walz}, \citenamefont
  {Berridge}, \citenamefont {Bugg}, \citenamefont {Shmakov},\ and\
  \citenamefont {Weidemann}}]{Bula_1996}%
  \BibitemOpen
  \bibfield  {author} {\bibinfo {author} {\bibfnamefont {C.}~\bibnamefont
  {Bula}}, \bibinfo {author} {\bibfnamefont {K.~T.}\ \bibnamefont {McDonald}},
  \bibinfo {author} {\bibfnamefont {E.~J.}\ \bibnamefont {Prebys}}, \bibinfo
  {author} {\bibfnamefont {C.}~\bibnamefont {Bamber}}, \bibinfo {author}
  {\bibfnamefont {S.}~\bibnamefont {Boege}}, \bibinfo {author} {\bibfnamefont
  {T.}~\bibnamefont {Kotseroglou}}, \bibinfo {author} {\bibfnamefont {A.~C.}\
  \bibnamefont {Melissinos}}, \bibinfo {author} {\bibfnamefont {D.~D.}\
  \bibnamefont {Meyerhofer}}, \bibinfo {author} {\bibfnamefont
  {W.}~\bibnamefont {Ragg}}, \bibinfo {author} {\bibfnamefont {D.~L.}\
  \bibnamefont {Burke}}, \bibinfo {author} {\bibfnamefont {R.~C.}\ \bibnamefont
  {Field}}, \bibinfo {author} {\bibfnamefont {G.}~\bibnamefont {Horton-Smith}},
  \bibinfo {author} {\bibfnamefont {A.~C.}\ \bibnamefont {Odian}}, \bibinfo
  {author} {\bibfnamefont {J.~E.}\ \bibnamefont {Spencer}}, \bibinfo {author}
  {\bibfnamefont {D.}~\bibnamefont {Walz}}, \bibinfo {author} {\bibfnamefont
  {S.~C.}\ \bibnamefont {Berridge}}, \bibinfo {author} {\bibfnamefont {W.~M.}\
  \bibnamefont {Bugg}}, \bibinfo {author} {\bibfnamefont {K.}~\bibnamefont
  {Shmakov}}, \ and\ \bibinfo {author} {\bibfnamefont {A.~W.}\ \bibnamefont
  {Weidemann}},\ }\href@noop {} {\bibfield  {journal} {\bibinfo  {journal}
  {Phys. Rev. Lett.}\ }\textbf {\bibinfo {volume} {76}},\ \bibinfo {pages}
  {3116} (\bibinfo {year} {1996})}\BibitemShut {NoStop}%
\bibitem [{\citenamefont {Burke}\ \emph {et~al.}(1997)\citenamefont {Burke},
  \citenamefont {Field}, \citenamefont {Horton-Smith}, \citenamefont {Spencer},
  \citenamefont {Walz}, \citenamefont {Berridge}, \citenamefont {Bugg},
  \citenamefont {Shmakov}, \citenamefont {Weidemann}, \citenamefont {Bula},
  \citenamefont {McDonald}, \citenamefont {Prebys}, \citenamefont {Bamber},
  \citenamefont {Boege}, \citenamefont {Koffas}, \citenamefont {Kotseroglou},
  \citenamefont {Melissinos}, \citenamefont {Meyerhofer}, \citenamefont
  {Reis},\ and\ \citenamefont {Ragg}}]{Burke_1997}%
  \BibitemOpen
  \bibfield  {author} {\bibinfo {author} {\bibfnamefont {D.~L.}\ \bibnamefont
  {Burke}}, \bibinfo {author} {\bibfnamefont {R.~C.}\ \bibnamefont {Field}},
  \bibinfo {author} {\bibfnamefont {G.}~\bibnamefont {Horton-Smith}}, \bibinfo
  {author} {\bibfnamefont {J.~E.}\ \bibnamefont {Spencer}}, \bibinfo {author}
  {\bibfnamefont {D.}~\bibnamefont {Walz}}, \bibinfo {author} {\bibfnamefont
  {S.~C.}\ \bibnamefont {Berridge}}, \bibinfo {author} {\bibfnamefont {W.~M.}\
  \bibnamefont {Bugg}}, \bibinfo {author} {\bibfnamefont {K.}~\bibnamefont
  {Shmakov}}, \bibinfo {author} {\bibfnamefont {A.~W.}\ \bibnamefont
  {Weidemann}}, \bibinfo {author} {\bibfnamefont {C.}~\bibnamefont {Bula}},
  \bibinfo {author} {\bibfnamefont {K.~T.}\ \bibnamefont {McDonald}}, \bibinfo
  {author} {\bibfnamefont {E.~J.}\ \bibnamefont {Prebys}}, \bibinfo {author}
  {\bibfnamefont {C.}~\bibnamefont {Bamber}}, \bibinfo {author} {\bibfnamefont
  {S.~J.}\ \bibnamefont {Boege}}, \bibinfo {author} {\bibfnamefont
  {T.}~\bibnamefont {Koffas}}, \bibinfo {author} {\bibfnamefont
  {T.}~\bibnamefont {Kotseroglou}}, \bibinfo {author} {\bibfnamefont {A.~C.}\
  \bibnamefont {Melissinos}}, \bibinfo {author} {\bibfnamefont {D.~D.}\
  \bibnamefont {Meyerhofer}}, \bibinfo {author} {\bibfnamefont {D.~A.}\
  \bibnamefont {Reis}}, \ and\ \bibinfo {author} {\bibfnamefont
  {W.}~\bibnamefont {Ragg}},\ }\href@noop {} {\bibfield  {journal} {\bibinfo
  {journal} {Phys. Rev. Lett.}\ }\textbf {\bibinfo {volume} {79}},\ \bibinfo
  {pages} {1626} (\bibinfo {year} {1997})}\BibitemShut {NoStop}%
\bibitem [{\citenamefont {Blackburn}\ \emph
  {et~al.}(2018{\natexlab{a}})\citenamefont {Blackburn}, \citenamefont
  {Ilderton}, \citenamefont {Marklund},\ and\ \citenamefont
  {Ridgers}}]{Blackburn_2018b}%
  \BibitemOpen
  \bibfield  {author} {\bibinfo {author} {\bibfnamefont {T.~G.}\ \bibnamefont
  {Blackburn}}, \bibinfo {author} {\bibfnamefont {A.}~\bibnamefont {Ilderton}},
  \bibinfo {author} {\bibfnamefont {M.}~\bibnamefont {Marklund}}, \ and\
  \bibinfo {author} {\bibfnamefont {C.~P.}\ \bibnamefont {Ridgers}},\
  }\href@noop {} {\bibfield  {journal} {\bibinfo  {journal} {arXiv:1807.03730}\
  } (\bibinfo {year} {2018}{\natexlab{a}})}\BibitemShut {NoStop}%
\bibitem [{\citenamefont {Baumann}\ \emph {et~al.}(2018)\citenamefont
  {Baumann}, \citenamefont {Nerush}, \citenamefont {Pukhov},\ and\
  \citenamefont {Kostyukov}}]{Baumann_2018}%
  \BibitemOpen
  \bibfield  {author} {\bibinfo {author} {\bibfnamefont {C.}~\bibnamefont
  {Baumann}}, \bibinfo {author} {\bibfnamefont {E.~N.}\ \bibnamefont {Nerush}},
  \bibinfo {author} {\bibfnamefont {A.}~\bibnamefont {Pukhov}}, \ and\ \bibinfo
  {author} {\bibfnamefont {I.~Y.}\ \bibnamefont {Kostyukov}},\ }\href@noop {}
  {\bibfield  {journal} {\bibinfo  {journal} {arXiv:1811.03990}\ } (\bibinfo
  {year} {2018})}\BibitemShut {NoStop}%
\bibitem [{\citenamefont {Yakimenko}\ \emph {et~al.}(2018)\citenamefont
  {Yakimenko}, \citenamefont {Meuren}, \citenamefont {Del~Gaudio},
  \citenamefont {Baumann}, \citenamefont {Fedotov}, \citenamefont {Fiuza},
  \citenamefont {Grismayer}, \citenamefont {Hogan}, \citenamefont {Pukhov},
  \citenamefont {Silva},\ and\ \citenamefont {White}}]{Yakimenko_2018}%
  \BibitemOpen
  \bibfield  {author} {\bibinfo {author} {\bibfnamefont {V.}~\bibnamefont
  {Yakimenko}}, \bibinfo {author} {\bibfnamefont {S.}~\bibnamefont {Meuren}},
  \bibinfo {author} {\bibfnamefont {F.}~\bibnamefont {Del~Gaudio}}, \bibinfo
  {author} {\bibfnamefont {C.}~\bibnamefont {Baumann}}, \bibinfo {author}
  {\bibfnamefont {A.~M.}\ \bibnamefont {Fedotov}}, \bibinfo {author}
  {\bibfnamefont {F.}~\bibnamefont {Fiuza}}, \bibinfo {author} {\bibfnamefont
  {T.}~\bibnamefont {Grismayer}}, \bibinfo {author} {\bibfnamefont {M.~J.}\
  \bibnamefont {Hogan}}, \bibinfo {author} {\bibfnamefont {A.}~\bibnamefont
  {Pukhov}}, \bibinfo {author} {\bibfnamefont {L.~O.}\ \bibnamefont {Silva}}, \
  and\ \bibinfo {author} {\bibfnamefont {G.}~\bibnamefont {White}},\
  }\href@noop {} {\bibfield  {journal} {\bibinfo  {journal} {arXiv:1807.09271}\
  } (\bibinfo {year} {2018})}\BibitemShut {NoStop}%
\bibitem [{\citenamefont {Ritus}(1970)}]{Ritus_1970}%
  \BibitemOpen
  \bibfield  {author} {\bibinfo {author} {\bibfnamefont {V.~I.}\ \bibnamefont
  {Ritus}},\ }\href@noop {} {\bibfield  {journal} {\bibinfo  {journal} {Sov.
  Phys. JETP}\ }\textbf {\bibinfo {volume} {30}},\ \bibinfo {pages} {1181}
  (\bibinfo {year} {1970})}\BibitemShut {NoStop}%
\bibitem [{\citenamefont {Narozhny}(1979)}]{Narozhny_1979}%
  \BibitemOpen
  \bibfield  {author} {\bibinfo {author} {\bibfnamefont {N.~B.}\ \bibnamefont
  {Narozhny}},\ }\href@noop {} {\bibfield  {journal} {\bibinfo  {journal}
  {Phys. Rev. D}\ }\textbf {\bibinfo {volume} {20}},\ \bibinfo {pages} {1313}
  (\bibinfo {year} {1979})}\BibitemShut {NoStop}%
\bibitem [{\citenamefont {Narozhny}(1980)}]{Narozhny_1980}%
  \BibitemOpen
  \bibfield  {author} {\bibinfo {author} {\bibfnamefont {N.~B.}\ \bibnamefont
  {Narozhny}},\ }\href@noop {} {\bibfield  {journal} {\bibinfo  {journal}
  {Phys. Rev. D}\ }\textbf {\bibinfo {volume} {21}},\ \bibinfo {pages} {1176}
  (\bibinfo {year} {1980})}\BibitemShut {NoStop}%
\bibitem [{\citenamefont {Morozov}\ \emph {et~al.}(1981)\citenamefont
  {Morozov}, \citenamefont {Narozhny},\ and\ \citenamefont
  {Ritus}}]{Morozov_1981}%
  \BibitemOpen
  \bibfield  {author} {\bibinfo {author} {\bibfnamefont {D.~A.}\ \bibnamefont
  {Morozov}}, \bibinfo {author} {\bibfnamefont {N.~B.}\ \bibnamefont
  {Narozhny}}, \ and\ \bibinfo {author} {\bibfnamefont {V.~I.}\ \bibnamefont
  {Ritus}},\ }\href@noop {} {\bibfield  {journal} {\bibinfo  {journal} {Sov.
  Phys. JETP}\ }\textbf {\bibinfo {volume} {53}},\ \bibinfo {pages} {1103}
  (\bibinfo {year} {1981})}\BibitemShut {NoStop}%
\bibitem [{\citenamefont {Akhmedov}(1983)}]{Akhmedov_1983}%
  \BibitemOpen
  \bibfield  {author} {\bibinfo {author} {\bibfnamefont {E.~K.}\ \bibnamefont
  {Akhmedov}},\ }\href@noop {} {\bibfield  {journal} {\bibinfo  {journal} {Sov.
  Phys. JETP}\ }\textbf {\bibinfo {volume} {58}},\ \bibinfo {pages} {883}
  (\bibinfo {year} {1983})}\BibitemShut {NoStop}%
\bibitem [{\citenamefont {Akhmedov}(2011)}]{Akhmedov_2011}%
  \BibitemOpen
  \bibfield  {author} {\bibinfo {author} {\bibfnamefont {E.~K.}\ \bibnamefont
  {Akhmedov}},\ }\href@noop {} {\bibfield  {journal} {\bibinfo  {journal}
  {Phys. Atom. Nucl.}\ }\textbf {\bibinfo {volume} {74}},\ \bibinfo {pages}
  {1299} (\bibinfo {year} {2011})}\BibitemShut {NoStop}%
\bibitem [{\citenamefont {Fedotov}(2017)}]{Fedotov_2017}%
  \BibitemOpen
  \bibfield  {author} {\bibinfo {author} {\bibfnamefont {A.~M.}\ \bibnamefont
  {Fedotov}},\ }\href@noop {} {\bibfield  {journal} {\bibinfo  {journal} {J.
  Phys.: Conf. Ser.}\ }\textbf {\bibinfo {volume} {826}},\ \bibinfo {pages}
  {012027} (\bibinfo {year} {2017})}\BibitemShut {NoStop}%
\bibitem [{\citenamefont {Becker}\ and\ \citenamefont
  {Mitter}(1975)}]{Becker_1975}%
  \BibitemOpen
  \bibfield  {author} {\bibinfo {author} {\bibfnamefont {W.}~\bibnamefont
  {Becker}}\ and\ \bibinfo {author} {\bibfnamefont {H.}~\bibnamefont
  {Mitter}},\ }\href@noop {} {\bibfield  {journal} {\bibinfo  {journal} {J.
  Phys. A}\ }\textbf {\bibinfo {volume} {8}},\ \bibinfo {pages} {1638}
  (\bibinfo {year} {1975})}\BibitemShut {NoStop}%
\bibitem [{\citenamefont {Baier}\ \emph
  {et~al.}(1976{\natexlab{a}})\citenamefont {Baier}, \citenamefont {Milstein},\
  and\ \citenamefont {Strakhovenko}}]{Baier_1976_b}%
  \BibitemOpen
  \bibfield  {author} {\bibinfo {author} {\bibfnamefont {V.~N.}\ \bibnamefont
  {Baier}}, \bibinfo {author} {\bibfnamefont {A.~I.}\ \bibnamefont {Milstein}},
  \ and\ \bibinfo {author} {\bibfnamefont {V.~M.}\ \bibnamefont
  {Strakhovenko}},\ }\href@noop {} {\bibfield  {journal} {\bibinfo  {journal}
  {Sov. Phys. JETP}\ }\textbf {\bibinfo {volume} {42}},\ \bibinfo {pages} {961}
  (\bibinfo {year} {1976}{\natexlab{a}})}\BibitemShut {NoStop}%
\bibitem [{\citenamefont {Meuren}\ \emph {et~al.}(2013)\citenamefont {Meuren},
  \citenamefont {Keitel},\ and\ \citenamefont {Di~Piazza}}]{Meuren_2013}%
  \BibitemOpen
  \bibfield  {author} {\bibinfo {author} {\bibfnamefont {S.}~\bibnamefont
  {Meuren}}, \bibinfo {author} {\bibfnamefont {C.~H.}\ \bibnamefont {Keitel}},
  \ and\ \bibinfo {author} {\bibfnamefont {A.}~\bibnamefont {Di~Piazza}},\
  }\href@noop {} {\bibfield  {journal} {\bibinfo  {journal} {Phys. Rev. D}\
  }\textbf {\bibinfo {volume} {88}},\ \bibinfo {pages} {013007} (\bibinfo
  {year} {2013})}\BibitemShut {NoStop}%
\bibitem [{\citenamefont {Baier}\ \emph
  {et~al.}(1976{\natexlab{b}})\citenamefont {Baier}, \citenamefont {Katkov},
  \citenamefont {Milstein},\ and\ \citenamefont {Strakhovenko}}]{Baier_1976_a}%
  \BibitemOpen
  \bibfield  {author} {\bibinfo {author} {\bibfnamefont {V.~N.}\ \bibnamefont
  {Baier}}, \bibinfo {author} {\bibfnamefont {V.~M.}\ \bibnamefont {Katkov}},
  \bibinfo {author} {\bibfnamefont {A.~I.}\ \bibnamefont {Milstein}}, \ and\
  \bibinfo {author} {\bibfnamefont {V.~M.}\ \bibnamefont {Strakhovenko}},\
  }\href@noop {} {\bibfield  {journal} {\bibinfo  {journal} {Sov. Phys. JETP}\
  }\textbf {\bibinfo {volume} {42}},\ \bibinfo {pages} {400} (\bibinfo {year}
  {1976}{\natexlab{b}})}\BibitemShut {NoStop}%
\bibitem [{\citenamefont {Furry}(1951)}]{Furry_1951}%
  \BibitemOpen
  \bibfield  {author} {\bibinfo {author} {\bibfnamefont {W.~H.}\ \bibnamefont
  {Furry}},\ }\href@noop {} {\bibfield  {journal} {\bibinfo  {journal} {Phys.
  Rev.}\ }\textbf {\bibinfo {volume} {81}},\ \bibinfo {pages} {115} (\bibinfo
  {year} {1951})}\BibitemShut {NoStop}%
\bibitem [{\citenamefont {Ilderton}(2019)}]{Ilderton_2019}%
  \BibitemOpen
  \bibfield  {author} {\bibinfo {author} {\bibfnamefont {A.}~\bibnamefont
  {Ilderton}},\ }\href@noop {} {\bibfield  {journal} {\bibinfo  {journal}
  {Phys. Rev. D}\ }\textbf {\bibinfo {volume} {99}},\ \bibinfo {pages} {085002}
  (\bibinfo {year} {2019})}\BibitemShut {NoStop}%
\bibitem [{\citenamefont {Ritus}(1972)}]{Ritus_1972}%
  \BibitemOpen
  \bibfield  {author} {\bibinfo {author} {\bibfnamefont {V.~I.}\ \bibnamefont
  {Ritus}},\ }\href@noop {} {\bibfield  {journal} {\bibinfo  {journal} {Ann.
  Phys. (N.Y.)}\ }\textbf {\bibinfo {volume} {69}},\ \bibinfo {pages} {555}
  (\bibinfo {year} {1972})}\BibitemShut {NoStop}%
\bibitem [{\citenamefont {Olver}\ \emph {et~al.}(2010)\citenamefont {Olver},
  \citenamefont {Lozier}, \citenamefont {Boisvert},\ and\ \citenamefont
  {Clark}}]{NIST_b_2010}%
  \BibitemOpen
  \bibinfo {editor} {\bibfnamefont {F.~W.~J.}\ \bibnamefont {Olver}}, \bibinfo
  {editor} {\bibfnamefont {D.~W.}\ \bibnamefont {Lozier}}, \bibinfo {editor}
  {\bibfnamefont {R.~F.}\ \bibnamefont {Boisvert}}, \ and\ \bibinfo {editor}
  {\bibfnamefont {C.~W.}\ \bibnamefont {Clark}},\ eds.,\ \href@noop {} {\emph
  {\bibinfo {title} {NIST Handbook of Mathematical Functions}}}\ (\bibinfo
  {publisher} {Cambridge University Press, Cambridge},\ \bibinfo {year}
  {2010})\BibitemShut {NoStop}%
\bibitem [{\citenamefont {Narozhny}(1969)}]{Narozhny_1969}%
  \BibitemOpen
  \bibfield  {author} {\bibinfo {author} {\bibfnamefont {N.~B.}\ \bibnamefont
  {Narozhny}},\ }\href@noop {} {\bibfield  {journal} {\bibinfo  {journal} {Sov.
  Phys. JETP}\ }\textbf {\bibinfo {volume} {28}},\ \bibinfo {pages} {371}
  (\bibinfo {year} {1969})}\BibitemShut {NoStop}%
\bibitem [{\citenamefont {Di~Piazza}\ \emph {et~al.}(2007)\citenamefont
  {Di~Piazza}, \citenamefont {Milstein},\ and\ \citenamefont
  {Keitel}}]{Di_Piazza_2007}%
  \BibitemOpen
  \bibfield  {author} {\bibinfo {author} {\bibfnamefont {A.}~\bibnamefont
  {Di~Piazza}}, \bibinfo {author} {\bibfnamefont {A.~I.}\ \bibnamefont
  {Milstein}}, \ and\ \bibinfo {author} {\bibfnamefont {C.~H.}\ \bibnamefont
  {Keitel}},\ }\href@noop {} {\bibfield  {journal} {\bibinfo  {journal} {Phys.
  Rev. A}\ }\textbf {\bibinfo {volume} {76}},\ \bibinfo {pages} {032103}
  (\bibinfo {year} {2007})}\BibitemShut {NoStop}%
\bibitem [{\citenamefont {Baier}\ \emph {et~al.}(1989)\citenamefont {Baier},
  \citenamefont {Katkov},\ and\ \citenamefont {Strakhovenko}}]{Baier_1989}%
  \BibitemOpen
  \bibfield  {author} {\bibinfo {author} {\bibfnamefont {V.~N.}\ \bibnamefont
  {Baier}}, \bibinfo {author} {\bibfnamefont {V.~M.}\ \bibnamefont {Katkov}}, \
  and\ \bibinfo {author} {\bibfnamefont {V.~M.}\ \bibnamefont {Strakhovenko}},\
  }\href@noop {} {\bibfield  {journal} {\bibinfo  {journal} {Nucl. Phys. B}\
  }\textbf {\bibinfo {volume} {328}},\ \bibinfo {pages} {387} (\bibinfo {year}
  {1989})}\BibitemShut {NoStop}%
\bibitem [{\citenamefont {Dinu}\ \emph {et~al.}(2016)\citenamefont {Dinu},
  \citenamefont {Harvey}, \citenamefont {Ilderton}, \citenamefont {Marklund},\
  and\ \citenamefont {Torgrimsson}}]{Dinu_2016}%
  \BibitemOpen
  \bibfield  {author} {\bibinfo {author} {\bibfnamefont {V.}~\bibnamefont
  {Dinu}}, \bibinfo {author} {\bibfnamefont {C.}~\bibnamefont {Harvey}},
  \bibinfo {author} {\bibfnamefont {A.}~\bibnamefont {Ilderton}}, \bibinfo
  {author} {\bibfnamefont {M.}~\bibnamefont {Marklund}}, \ and\ \bibinfo
  {author} {\bibfnamefont {G.}~\bibnamefont {Torgrimsson}},\ }\href@noop {}
  {\bibfield  {journal} {\bibinfo  {journal} {Phys. Rev. Lett.}\ }\textbf
  {\bibinfo {volume} {116}},\ \bibinfo {pages} {044801} (\bibinfo {year}
  {2016})}\BibitemShut {NoStop}%
\bibitem [{\citenamefont {Di~Piazza}\ \emph {et~al.}(2018)\citenamefont
  {Di~Piazza}, \citenamefont {Tamburini}, \citenamefont {Meuren},\ and\
  \citenamefont {Keitel}}]{Di_Piazza_2018c}%
  \BibitemOpen
  \bibfield  {author} {\bibinfo {author} {\bibfnamefont {A.}~\bibnamefont
  {Di~Piazza}}, \bibinfo {author} {\bibfnamefont {M.}~\bibnamefont
  {Tamburini}}, \bibinfo {author} {\bibfnamefont {S.}~\bibnamefont {Meuren}}, \
  and\ \bibinfo {author} {\bibfnamefont {C.~H.}\ \bibnamefont {Keitel}},\
  }\href@noop {} {\bibfield  {journal} {\bibinfo  {journal} {Phys. Rev. A}\
  }\textbf {\bibinfo {volume} {98}},\ \bibinfo {pages} {012134} (\bibinfo
  {year} {2018})}\BibitemShut {NoStop}%
\bibitem [{\citenamefont {Gradshteyn}\ and\ \citenamefont
  {Ryzhik}(2000)}]{Gradshteyn_b_2000}%
  \BibitemOpen
  \bibfield  {author} {\bibinfo {author} {\bibfnamefont {I.~S.}\ \bibnamefont
  {Gradshteyn}}\ and\ \bibinfo {author} {\bibfnamefont {I.~M.}\ \bibnamefont
  {Ryzhik}},\ }\href@noop {} {\emph {\bibinfo {title} {Tables of Integrals,
  Series and Products}}}\ (\bibinfo  {publisher} {Academic Press, San Diego},\
  \bibinfo {year} {2000})\BibitemShut {NoStop}%
\bibitem [{\citenamefont {Mackenroth}\ and\ \citenamefont
  {Di~Piazza}(2011)}]{Mackenroth_2011}%
  \BibitemOpen
  \bibfield  {author} {\bibinfo {author} {\bibfnamefont {F.}~\bibnamefont
  {Mackenroth}}\ and\ \bibinfo {author} {\bibfnamefont {A.}~\bibnamefont
  {Di~Piazza}},\ }\href@noop {} {\bibfield  {journal} {\bibinfo  {journal}
  {Phys. Rev. A}\ }\textbf {\bibinfo {volume} {83}},\ \bibinfo {pages} {032106}
  (\bibinfo {year} {2011})}\BibitemShut {NoStop}%
\bibitem [{\citenamefont {Bender}\ and\ \citenamefont
  {Orszag}(1999)}]{Bender_b_1999}%
  \BibitemOpen
  \bibfield  {author} {\bibinfo {author} {\bibfnamefont {C.~M.}\ \bibnamefont
  {Bender}}\ and\ \bibinfo {author} {\bibfnamefont {S.~A.}\ \bibnamefont
  {Orszag}},\ }\href@noop {} {\emph {\bibinfo {title} {Advanced Mathematical
  Methods for Scientists and Engineers}}}\ (\bibinfo  {publisher}
  {Springer-Verlag, New York},\ \bibinfo {year} {1999})\BibitemShut {NoStop}%
\bibitem [{\citenamefont {Reiss}(1962)}]{Reiss_1962}%
  \BibitemOpen
  \bibfield  {author} {\bibinfo {author} {\bibfnamefont {H.~R.}\ \bibnamefont
  {Reiss}},\ }\href@noop {} {\bibfield  {journal} {\bibinfo  {journal} {J.
  Math. Phys. (N.Y.)}\ }\textbf {\bibinfo {volume} {3}},\ \bibinfo {pages} {59}
  (\bibinfo {year} {1962})}\BibitemShut {NoStop}%
\bibitem [{\citenamefont {Nikishov}\ and\ \citenamefont
  {Ritus}(1964)}]{Nikishov_1964}%
  \BibitemOpen
  \bibfield  {author} {\bibinfo {author} {\bibfnamefont {A.~I.}\ \bibnamefont
  {Nikishov}}\ and\ \bibinfo {author} {\bibfnamefont {V.~I.}\ \bibnamefont
  {Ritus}},\ }\href@noop {} {\bibfield  {journal} {\bibinfo  {journal} {Sov.
  Phys. JETP}\ }\textbf {\bibinfo {volume} {19}},\ \bibinfo {pages} {529}
  (\bibinfo {year} {1964})}\BibitemShut {NoStop}%
\bibitem [{\citenamefont {Narozhny}\ and\ \citenamefont
  {Fofanov}(2000)}]{Narozhny_2000}%
  \BibitemOpen
  \bibfield  {author} {\bibinfo {author} {\bibfnamefont {N.~B.}\ \bibnamefont
  {Narozhny}}\ and\ \bibinfo {author} {\bibfnamefont {M.~S.}\ \bibnamefont
  {Fofanov}},\ }\href@noop {} {\bibfield  {journal} {\bibinfo  {journal} {J.
  Exp. Theor. Phys.}\ }\textbf {\bibinfo {volume} {90}},\ \bibinfo {pages}
  {415} (\bibinfo {year} {2000})}\BibitemShut {NoStop}%
\bibitem [{\citenamefont {Roshchupkin}(2001)}]{Roshchupkin_2001}%
  \BibitemOpen
  \bibfield  {author} {\bibinfo {author} {\bibfnamefont {S.~P.}\ \bibnamefont
  {Roshchupkin}},\ }\href@noop {} {\bibfield  {journal} {\bibinfo  {journal}
  {Phys. At. Nucl.}\ }\textbf {\bibinfo {volume} {64}},\ \bibinfo {pages} {243}
  (\bibinfo {year} {2001})}\BibitemShut {NoStop}%
\bibitem [{\citenamefont {Heinzl}\ \emph {et~al.}(2010)\citenamefont {Heinzl},
  \citenamefont {Ilderton},\ and\ \citenamefont {Marklund}}]{Heinzl_2010b}%
  \BibitemOpen
  \bibfield  {author} {\bibinfo {author} {\bibfnamefont {T.}~\bibnamefont
  {Heinzl}}, \bibinfo {author} {\bibfnamefont {A.}~\bibnamefont {Ilderton}}, \
  and\ \bibinfo {author} {\bibfnamefont {M.}~\bibnamefont {Marklund}},\
  }\href@noop {} {\bibfield  {journal} {\bibinfo  {journal} {Phys. Lett. B}\
  }\textbf {\bibinfo {volume} {692}},\ \bibinfo {pages} {250} (\bibinfo {year}
  {2010})}\BibitemShut {NoStop}%
\bibitem [{\citenamefont {M\"{u}ller}\ and\ \citenamefont
  {M\"{u}ller}(2011)}]{Mueller_2011b}%
  \BibitemOpen
  \bibfield  {author} {\bibinfo {author} {\bibfnamefont {T.-O.}\ \bibnamefont
  {M\"{u}ller}}\ and\ \bibinfo {author} {\bibfnamefont {C.}~\bibnamefont
  {M\"{u}ller}},\ }\href@noop {} {\bibfield  {journal} {\bibinfo  {journal}
  {Phys. Lett. B}\ }\textbf {\bibinfo {volume} {696}},\ \bibinfo {pages} {201}
  (\bibinfo {year} {2011})}\BibitemShut {NoStop}%
\bibitem [{\citenamefont {Titov}\ \emph {et~al.}(2012)\citenamefont {Titov},
  \citenamefont {Takabe}, \citenamefont {K\"ampfer},\ and\ \citenamefont
  {Hosaka}}]{Titov_2012}%
  \BibitemOpen
  \bibfield  {author} {\bibinfo {author} {\bibfnamefont {A.~I.}\ \bibnamefont
  {Titov}}, \bibinfo {author} {\bibfnamefont {H.}~\bibnamefont {Takabe}},
  \bibinfo {author} {\bibfnamefont {B.}~\bibnamefont {K\"ampfer}}, \ and\
  \bibinfo {author} {\bibfnamefont {A.}~\bibnamefont {Hosaka}},\ }\href@noop {}
  {\bibfield  {journal} {\bibinfo  {journal} {Phys. Rev. Lett.}\ }\textbf
  {\bibinfo {volume} {108}},\ \bibinfo {pages} {240406} (\bibinfo {year}
  {2012})}\BibitemShut {NoStop}%
\bibitem [{\citenamefont {Nousch}\ \emph {et~al.}(2012)\citenamefont {Nousch},
  \citenamefont {Seipt}, \citenamefont {K\"{a}mpfer},\ and\ \citenamefont
  {Titov}}]{Nousch_2012}%
  \BibitemOpen
  \bibfield  {author} {\bibinfo {author} {\bibfnamefont {T.}~\bibnamefont
  {Nousch}}, \bibinfo {author} {\bibfnamefont {D.}~\bibnamefont {Seipt}},
  \bibinfo {author} {\bibfnamefont {B.}~\bibnamefont {K\"{a}mpfer}}, \ and\
  \bibinfo {author} {\bibfnamefont {A.}~\bibnamefont {Titov}},\ }\href@noop {}
  {\bibfield  {journal} {\bibinfo  {journal} {Phys. Lett. B}\ }\textbf
  {\bibinfo {volume} {715}},\ \bibinfo {pages} {246} (\bibinfo {year}
  {2012})}\BibitemShut {NoStop}%
\bibitem [{\citenamefont {Krajewska}\ \emph {et~al.}(2013)\citenamefont
  {Krajewska}, \citenamefont {M\"uller},\ and\ \citenamefont
  {Kami\ifmmode~\acute{n}\else \'{n}\fi{}ski}}]{Krajewska_2013b}%
  \BibitemOpen
  \bibfield  {author} {\bibinfo {author} {\bibfnamefont {K.}~\bibnamefont
  {Krajewska}}, \bibinfo {author} {\bibfnamefont {C.}~\bibnamefont {M\"uller}},
  \ and\ \bibinfo {author} {\bibfnamefont {J.~Z.}\ \bibnamefont
  {Kami\ifmmode~\acute{n}\else \'{n}\fi{}ski}},\ }\href@noop {} {\bibfield
  {journal} {\bibinfo  {journal} {Phys. Rev. A}\ }\textbf {\bibinfo {volume}
  {87}},\ \bibinfo {pages} {062107} (\bibinfo {year} {2013})}\BibitemShut
  {NoStop}%
\bibitem [{\citenamefont {Jansen}\ and\ \citenamefont
  {M\"uller}(2013)}]{Jansen_2013}%
  \BibitemOpen
  \bibfield  {author} {\bibinfo {author} {\bibfnamefont {M.~J.~A.}\
  \bibnamefont {Jansen}}\ and\ \bibinfo {author} {\bibfnamefont
  {C.}~\bibnamefont {M\"uller}},\ }\href@noop {} {\bibfield  {journal}
  {\bibinfo  {journal} {Phys. Rev. A}\ }\textbf {\bibinfo {volume} {88}},\
  \bibinfo {pages} {052125} (\bibinfo {year} {2013})}\BibitemShut {NoStop}%
\bibitem [{\citenamefont {Augustin}\ and\ \citenamefont
  {M\"{u}ller}(2014)}]{Augustin_2014}%
  \BibitemOpen
  \bibfield  {author} {\bibinfo {author} {\bibfnamefont {S.}~\bibnamefont
  {Augustin}}\ and\ \bibinfo {author} {\bibfnamefont {C.}~\bibnamefont
  {M\"{u}ller}},\ }\href@noop {} {\bibfield  {journal} {\bibinfo  {journal}
  {Phys. Lett. B}\ }\textbf {\bibinfo {volume} {737}},\ \bibinfo {pages} {114}
  (\bibinfo {year} {2014})}\BibitemShut {NoStop}%
\bibitem [{\citenamefont {Meuren}\ \emph {et~al.}(2015)\citenamefont {Meuren},
  \citenamefont {Hatsagortsyan}, \citenamefont {Keitel},\ and\ \citenamefont
  {Di~Piazza}}]{Meuren_2015}%
  \BibitemOpen
  \bibfield  {author} {\bibinfo {author} {\bibfnamefont {S.}~\bibnamefont
  {Meuren}}, \bibinfo {author} {\bibfnamefont {K.~Z.}\ \bibnamefont
  {Hatsagortsyan}}, \bibinfo {author} {\bibfnamefont {C.~H.}\ \bibnamefont
  {Keitel}}, \ and\ \bibinfo {author} {\bibfnamefont {A.}~\bibnamefont
  {Di~Piazza}},\ }\href@noop {} {\bibfield  {journal} {\bibinfo  {journal}
  {Phys. Rev. D}\ }\textbf {\bibinfo {volume} {91}},\ \bibinfo {pages} {013009}
  (\bibinfo {year} {2015})}\BibitemShut {NoStop}%
\bibitem [{\citenamefont {Meuren}\ \emph {et~al.}(2016)\citenamefont {Meuren},
  \citenamefont {Keitel},\ and\ \citenamefont {Di~Piazza}}]{Meuren_2016}%
  \BibitemOpen
  \bibfield  {author} {\bibinfo {author} {\bibfnamefont {S.}~\bibnamefont
  {Meuren}}, \bibinfo {author} {\bibfnamefont {C.~H.}\ \bibnamefont {Keitel}},
  \ and\ \bibinfo {author} {\bibfnamefont {A.}~\bibnamefont {Di~Piazza}},\
  }\href@noop {} {\bibfield  {journal} {\bibinfo  {journal} {Phys. Rev. D}\
  }\textbf {\bibinfo {volume} {93}},\ \bibinfo {pages} {085028} (\bibinfo
  {year} {2016})}\BibitemShut {NoStop}%
\bibitem [{\citenamefont {Blackburn}\ and\ \citenamefont
  {Marklund}(2018)}]{Blackburn_2018c}%
  \BibitemOpen
  \bibfield  {author} {\bibinfo {author} {\bibfnamefont {T.~G.}\ \bibnamefont
  {Blackburn}}\ and\ \bibinfo {author} {\bibfnamefont {M.}~\bibnamefont
  {Marklund}},\ }\href@noop {} {\bibfield  {journal} {\bibinfo  {journal}
  {Plasma Phys. Control. Fusion}\ }\textbf {\bibinfo {volume} {60}},\ \bibinfo
  {pages} {054009} (\bibinfo {year} {2018})}\BibitemShut {NoStop}%
\bibitem [{\citenamefont {Di~Piazza}\ \emph {et~al.}(2019)\citenamefont
  {Di~Piazza}, \citenamefont {Tamburini}, \citenamefont {Meuren},\ and\
  \citenamefont {Keitel}}]{Di_Piazza_2019d}%
  \BibitemOpen
  \bibfield  {author} {\bibinfo {author} {\bibfnamefont {A.}~\bibnamefont
  {Di~Piazza}}, \bibinfo {author} {\bibfnamefont {M.}~\bibnamefont
  {Tamburini}}, \bibinfo {author} {\bibfnamefont {S.}~\bibnamefont {Meuren}}, \
  and\ \bibinfo {author} {\bibfnamefont {C.~H.}\ \bibnamefont {Keitel}},\
  }\href@noop {} {\bibfield  {journal} {\bibinfo  {journal} {Phys. Rev. A}\
  }\textbf {\bibinfo {volume} {99}},\ \bibinfo {pages} {022125} (\bibinfo
  {year} {2019})}\BibitemShut {NoStop}%
\bibitem [{\citenamefont {Ivanov}\ \emph {et~al.}(2004)\citenamefont {Ivanov},
  \citenamefont {Kotkin},\ and\ \citenamefont {Serbo}}]{Ivanov_2004}%
  \BibitemOpen
  \bibfield  {author} {\bibinfo {author} {\bibfnamefont {D.~{\relax Yu}.}\
  \bibnamefont {Ivanov}}, \bibinfo {author} {\bibfnamefont {G.~L.}\
  \bibnamefont {Kotkin}}, \ and\ \bibinfo {author} {\bibfnamefont {V.~G.}\
  \bibnamefont {Serbo}},\ }\href@noop {} {\bibfield  {journal} {\bibinfo
  {journal} {Eur. Phys. J. C}\ }\textbf {\bibinfo {volume} {36}},\ \bibinfo
  {pages} {127} (\bibinfo {year} {2004})}\BibitemShut {NoStop}%
\bibitem [{\citenamefont {Boca}\ and\ \citenamefont
  {Florescu}(2009)}]{Boca_2009}%
  \BibitemOpen
  \bibfield  {author} {\bibinfo {author} {\bibfnamefont {M.}~\bibnamefont
  {Boca}}\ and\ \bibinfo {author} {\bibfnamefont {V.}~\bibnamefont
  {Florescu}},\ }\href@noop {} {\bibfield  {journal} {\bibinfo  {journal}
  {Phys. Rev. A}\ }\textbf {\bibinfo {volume} {80}},\ \bibinfo {pages} {053403}
  (\bibinfo {year} {2009})}\BibitemShut {NoStop}%
\bibitem [{\citenamefont {Harvey}\ \emph {et~al.}(2009)\citenamefont {Harvey},
  \citenamefont {Heinzl},\ and\ \citenamefont {Ilderton}}]{Harvey_2009}%
  \BibitemOpen
  \bibfield  {author} {\bibinfo {author} {\bibfnamefont {C.}~\bibnamefont
  {Harvey}}, \bibinfo {author} {\bibfnamefont {T.}~\bibnamefont {Heinzl}}, \
  and\ \bibinfo {author} {\bibfnamefont {A.}~\bibnamefont {Ilderton}},\
  }\href@noop {} {\bibfield  {journal} {\bibinfo  {journal} {Phys. Rev. A}\
  }\textbf {\bibinfo {volume} {79}},\ \bibinfo {pages} {063407} (\bibinfo
  {year} {2009})}\BibitemShut {NoStop}%
\bibitem [{\citenamefont {Mackenroth}\ \emph {et~al.}(2010)\citenamefont
  {Mackenroth}, \citenamefont {Di~Piazza},\ and\ \citenamefont
  {Keitel}}]{Mackenroth_2010}%
  \BibitemOpen
  \bibfield  {author} {\bibinfo {author} {\bibfnamefont {F.}~\bibnamefont
  {Mackenroth}}, \bibinfo {author} {\bibfnamefont {A.}~\bibnamefont
  {Di~Piazza}}, \ and\ \bibinfo {author} {\bibfnamefont {C.~H.}\ \bibnamefont
  {Keitel}},\ }\href@noop {} {\bibfield  {journal} {\bibinfo  {journal} {Phys.
  Rev. Lett.}\ }\textbf {\bibinfo {volume} {105}},\ \bibinfo {pages} {063903}
  (\bibinfo {year} {2010})}\BibitemShut {NoStop}%
\bibitem [{\citenamefont {Boca}\ and\ \citenamefont
  {Florescu}(2011)}]{Boca_2011}%
  \BibitemOpen
  \bibfield  {author} {\bibinfo {author} {\bibfnamefont {M.}~\bibnamefont
  {Boca}}\ and\ \bibinfo {author} {\bibfnamefont {V.}~\bibnamefont
  {Florescu}},\ }\href@noop {} {\bibfield  {journal} {\bibinfo  {journal} {Eur.
  Phys. J. D}\ }\textbf {\bibinfo {volume} {61}},\ \bibinfo {pages} {449}
  (\bibinfo {year} {2011})}\BibitemShut {NoStop}%
\bibitem [{\citenamefont {Seipt}\ and\ \citenamefont
  {K\"ampfer}(2011{\natexlab{a}})}]{Seipt_2011}%
  \BibitemOpen
  \bibfield  {author} {\bibinfo {author} {\bibfnamefont {D.}~\bibnamefont
  {Seipt}}\ and\ \bibinfo {author} {\bibfnamefont {B.}~\bibnamefont
  {K\"ampfer}},\ }\href@noop {} {\bibfield  {journal} {\bibinfo  {journal}
  {Phys. Rev. A}\ }\textbf {\bibinfo {volume} {83}},\ \bibinfo {pages} {022101}
  (\bibinfo {year} {2011}{\natexlab{a}})}\BibitemShut {NoStop}%
\bibitem [{\citenamefont {Seipt}\ and\ \citenamefont
  {K\"ampfer}(2011{\natexlab{b}})}]{Seipt_2011b}%
  \BibitemOpen
  \bibfield  {author} {\bibinfo {author} {\bibfnamefont {D.}~\bibnamefont
  {Seipt}}\ and\ \bibinfo {author} {\bibfnamefont {B.}~\bibnamefont
  {K\"ampfer}},\ }\href@noop {} {\bibfield  {journal} {\bibinfo  {journal}
  {Phys. Rev. Accel. Beams}\ }\textbf {\bibinfo {volume} {14}},\ \bibinfo
  {pages} {040704} (\bibinfo {year} {2011}{\natexlab{b}})}\BibitemShut
  {NoStop}%
\bibitem [{\citenamefont {Dinu}\ \emph {et~al.}(2012)\citenamefont {Dinu},
  \citenamefont {Heinzl},\ and\ \citenamefont {Ilderton}}]{Dinu_2012}%
  \BibitemOpen
  \bibfield  {author} {\bibinfo {author} {\bibfnamefont {V.}~\bibnamefont
  {Dinu}}, \bibinfo {author} {\bibfnamefont {T.}~\bibnamefont {Heinzl}}, \ and\
  \bibinfo {author} {\bibfnamefont {A.}~\bibnamefont {Ilderton}},\ }\href@noop
  {} {\bibfield  {journal} {\bibinfo  {journal} {Phys. Rev. D}\ }\textbf
  {\bibinfo {volume} {86}},\ \bibinfo {pages} {085037} (\bibinfo {year}
  {2012})}\BibitemShut {NoStop}%
\bibitem [{\citenamefont {Krajewska}\ and\ \citenamefont
  {Kami\ifmmode~\acute{n}\else \'{n}\fi{}ski}(2012)}]{Krajewska_2012}%
  \BibitemOpen
  \bibfield  {author} {\bibinfo {author} {\bibfnamefont {K.}~\bibnamefont
  {Krajewska}}\ and\ \bibinfo {author} {\bibfnamefont {J.~Z.}\ \bibnamefont
  {Kami\ifmmode~\acute{n}\else \'{n}\fi{}ski}},\ }\href@noop {} {\bibfield
  {journal} {\bibinfo  {journal} {Phys. Rev. A}\ }\textbf {\bibinfo {volume}
  {85}},\ \bibinfo {pages} {062102} (\bibinfo {year} {2012})}\BibitemShut
  {NoStop}%
\bibitem [{\citenamefont {Dinu}(2013)}]{Dinu_2013}%
  \BibitemOpen
  \bibfield  {author} {\bibinfo {author} {\bibfnamefont {V.}~\bibnamefont
  {Dinu}},\ }\href@noop {} {\bibfield  {journal} {\bibinfo  {journal} {Phys.
  Rev. A}\ }\textbf {\bibinfo {volume} {87}},\ \bibinfo {pages} {052101}
  (\bibinfo {year} {2013})}\BibitemShut {NoStop}%
\bibitem [{\citenamefont {Seipt}\ and\ \citenamefont
  {K\"ampfer}(2013)}]{Seipt_2013}%
  \BibitemOpen
  \bibfield  {author} {\bibinfo {author} {\bibfnamefont {D.}~\bibnamefont
  {Seipt}}\ and\ \bibinfo {author} {\bibfnamefont {B.}~\bibnamefont
  {K\"ampfer}},\ }\href@noop {} {\bibfield  {journal} {\bibinfo  {journal}
  {Phys. Rev. A}\ }\textbf {\bibinfo {volume} {88}},\ \bibinfo {pages} {012127}
  (\bibinfo {year} {2013})}\BibitemShut {NoStop}%
\bibitem [{\citenamefont {Krajewska}\ \emph {et~al.}(2014)\citenamefont
  {Krajewska}, \citenamefont {Twardy},\ and\ \citenamefont
  {Kami\ifmmode~\acute{n}\else \'{n}\fi{}ski}}]{Krajewska_2014}%
  \BibitemOpen
  \bibfield  {author} {\bibinfo {author} {\bibfnamefont {K.}~\bibnamefont
  {Krajewska}}, \bibinfo {author} {\bibfnamefont {M.}~\bibnamefont {Twardy}}, \
  and\ \bibinfo {author} {\bibfnamefont {J.~Z.}\ \bibnamefont
  {Kami\ifmmode~\acute{n}\else \'{n}\fi{}ski}},\ }\href@noop {} {\bibfield
  {journal} {\bibinfo  {journal} {Phys. Rev. A}\ }\textbf {\bibinfo {volume}
  {89}},\ \bibinfo {pages} {032125} (\bibinfo {year} {2014})}\BibitemShut
  {NoStop}%
\bibitem [{\citenamefont {Wistisen}(2014)}]{Wistisen_2014}%
  \BibitemOpen
  \bibfield  {author} {\bibinfo {author} {\bibfnamefont {T.~N.}\ \bibnamefont
  {Wistisen}},\ }\href@noop {} {\bibfield  {journal} {\bibinfo  {journal}
  {Phys. Rev. D}\ }\textbf {\bibinfo {volume} {90}},\ \bibinfo {pages} {125008}
  (\bibinfo {year} {2014})}\BibitemShut {NoStop}%
\bibitem [{\citenamefont {Harvey}\ \emph {et~al.}(2015)\citenamefont {Harvey},
  \citenamefont {Ilderton},\ and\ \citenamefont {King}}]{Harvey_2015}%
  \BibitemOpen
  \bibfield  {author} {\bibinfo {author} {\bibfnamefont {C.~N.}\ \bibnamefont
  {Harvey}}, \bibinfo {author} {\bibfnamefont {A.}~\bibnamefont {Ilderton}}, \
  and\ \bibinfo {author} {\bibfnamefont {B.}~\bibnamefont {King}},\ }\href@noop
  {} {\bibfield  {journal} {\bibinfo  {journal} {Phys. Rev. A}\ }\textbf
  {\bibinfo {volume} {91}},\ \bibinfo {pages} {013822} (\bibinfo {year}
  {2015})}\BibitemShut {NoStop}%
\bibitem [{\citenamefont {Seipt}\ \emph
  {et~al.}(2016{\natexlab{a}})\citenamefont {Seipt}, \citenamefont {Kharin},
  \citenamefont {Rykovanov}, \citenamefont {Surzhykov},\ and\ \citenamefont
  {Fritzsche}}]{Seipt_2016}%
  \BibitemOpen
  \bibfield  {author} {\bibinfo {author} {\bibfnamefont {D.}~\bibnamefont
  {Seipt}}, \bibinfo {author} {\bibfnamefont {V.}~\bibnamefont {Kharin}},
  \bibinfo {author} {\bibfnamefont {S.}~\bibnamefont {Rykovanov}}, \bibinfo
  {author} {\bibfnamefont {A.}~\bibnamefont {Surzhykov}}, \ and\ \bibinfo
  {author} {\bibfnamefont {S.}~\bibnamefont {Fritzsche}},\ }\href@noop {}
  {\bibfield  {journal} {\bibinfo  {journal} {J. Plasma Phys.}\ }\textbf
  {\bibinfo {volume} {82}},\ \bibinfo {pages} {655820203} (\bibinfo {year}
  {2016}{\natexlab{a}})}\BibitemShut {NoStop}%
\bibitem [{\citenamefont {Seipt}\ \emph
  {et~al.}(2016{\natexlab{b}})\citenamefont {Seipt}, \citenamefont {Surzhykov},
  \citenamefont {Fritzsche},\ and\ \citenamefont {K\"{a}mpfer}}]{Seipt_2016b}%
  \BibitemOpen
  \bibfield  {author} {\bibinfo {author} {\bibfnamefont {D.}~\bibnamefont
  {Seipt}}, \bibinfo {author} {\bibfnamefont {A.}~\bibnamefont {Surzhykov}},
  \bibinfo {author} {\bibfnamefont {S.}~\bibnamefont {Fritzsche}}, \ and\
  \bibinfo {author} {\bibfnamefont {B.}~\bibnamefont {K\"{a}mpfer}},\
  }\href@noop {} {\bibfield  {journal} {\bibinfo  {journal} {New J. Phys.}\
  }\textbf {\bibinfo {volume} {18}},\ \bibinfo {pages} {023044} (\bibinfo
  {year} {2016}{\natexlab{b}})}\BibitemShut {NoStop}%
\bibitem [{\citenamefont {Angioi}\ \emph {et~al.}(2016)\citenamefont {Angioi},
  \citenamefont {Mackenroth},\ and\ \citenamefont {Di~Piazza}}]{Angioi_2016}%
  \BibitemOpen
  \bibfield  {author} {\bibinfo {author} {\bibfnamefont {A.}~\bibnamefont
  {Angioi}}, \bibinfo {author} {\bibfnamefont {F.}~\bibnamefont {Mackenroth}},
  \ and\ \bibinfo {author} {\bibfnamefont {A.}~\bibnamefont {Di~Piazza}},\
  }\href@noop {} {\bibfield  {journal} {\bibinfo  {journal} {Phys. Rev. A}\
  }\textbf {\bibinfo {volume} {93}},\ \bibinfo {pages} {052102} (\bibinfo
  {year} {2016})}\BibitemShut {NoStop}%
\bibitem [{\citenamefont {Harvey}\ \emph {et~al.}(2016)\citenamefont {Harvey},
  \citenamefont {Gonoskov}, \citenamefont {Marklund},\ and\ \citenamefont
  {Wallin}}]{Harvey_2016b}%
  \BibitemOpen
  \bibfield  {author} {\bibinfo {author} {\bibfnamefont {C.~N.}\ \bibnamefont
  {Harvey}}, \bibinfo {author} {\bibfnamefont {A.}~\bibnamefont {Gonoskov}},
  \bibinfo {author} {\bibfnamefont {M.}~\bibnamefont {Marklund}}, \ and\
  \bibinfo {author} {\bibfnamefont {E.}~\bibnamefont {Wallin}},\ }\href@noop {}
  {\bibfield  {journal} {\bibinfo  {journal} {Phys. Rev. A}\ }\textbf {\bibinfo
  {volume} {93}},\ \bibinfo {pages} {022112} (\bibinfo {year}
  {2016})}\BibitemShut {NoStop}%
\bibitem [{\citenamefont {Ilderton}\ \emph {et~al.}(2018)\citenamefont
  {Ilderton}, \citenamefont {King},\ and\ \citenamefont
  {Seipt}}]{Ilderton_2018b}%
  \BibitemOpen
  \bibfield  {author} {\bibinfo {author} {\bibfnamefont {A.}~\bibnamefont
  {Ilderton}}, \bibinfo {author} {\bibfnamefont {B.}~\bibnamefont {King}}, \
  and\ \bibinfo {author} {\bibfnamefont {D.}~\bibnamefont {Seipt}},\
  }\href@noop {} {\bibfield  {journal} {\bibinfo  {journal} {arXiv:1808.10339}\
  } (\bibinfo {year} {2018})}\BibitemShut {NoStop}%
\bibitem [{\citenamefont {Blackburn}\ \emph
  {et~al.}(2018{\natexlab{b}})\citenamefont {Blackburn}, \citenamefont {Seipt},
  \citenamefont {Bulanov},\ and\ \citenamefont {Marklund}}]{Blackburn_2018}%
  \BibitemOpen
  \bibfield  {author} {\bibinfo {author} {\bibfnamefont {T.~G.}\ \bibnamefont
  {Blackburn}}, \bibinfo {author} {\bibfnamefont {D.}~\bibnamefont {Seipt}},
  \bibinfo {author} {\bibfnamefont {S.~S.}\ \bibnamefont {Bulanov}}, \ and\
  \bibinfo {author} {\bibfnamefont {M.}~\bibnamefont {Marklund}},\ }\href@noop
  {} {\bibfield  {journal} {\bibinfo  {journal} {Phys. Plasmas}\ }\textbf
  {\bibinfo {volume} {25}},\ \bibinfo {pages} {083108} (\bibinfo {year}
  {2018}{\natexlab{b}})}\BibitemShut {NoStop}%
\end{thebibliography}
\end{document}